# Imprinting the quantum statistics of photons on free electrons

Raphael Dahan[1†], Alexey Gorlach[1†], Urs Haeusler[2,3†], Aviv Karnieli[1,4†], Ori Eyal[1], Peyman Yousefi[2,5], Mordechai Segev[1,6], Ady Arie[7], Gadi Eisenstein[1], Peter Hommelhoff[2], and Ido Kaminer[1*]

[1] Department of Electrical Engineering, Russell Berrie Nanotechnology Institute, and Solid-State Institute, Technion – Israel Institute of Technology, Haifa 32000, Israel

[2] Department of Physics, Friedrich-Alexander-Universität Erlangen-Nürnberg (FAU), Staudtstraße 1, Erlangen 91058, Germany

[3] Now at Cavendish Laboratory, University of Cambridge, JJ Thomson Avenue, Cambridge CB3 0HE, UK

[4] Raymond and Beverly Sackler School of Physics and Astronomy, Tel Aviv University, Tel Aviv 69978, Israel

[5] Now at Fraunhofer-Institut für Keramische Technologien und Systeme IKTS, Äussere Nürnberger Strasse 62, Forchheim 91301, Germany

[6] Department of Physics, Technion – Israel Institute of Technology, Haifa 32000, Israel

[7] School of Electrical Engineering, Fleischman Faculty of Engineering, Tel Aviv University, Tel Aviv 69978, Israel

[†] *equal contributors*

\* *kaminer@technion.ac.il*

**Abstract**

The fundamental interaction between free electrons and light stands at the base of both classical and quantum physics, with applications in free-electron acceleration, radiation sources, and electron microscopy. Yet, to this day, all experiments involving free-electron–light interactions are fully explained by describing the light as a classical wave, disregarding its quantum nature. Here, we observe quantum statistics effects of photons on free-electron–light interactions. We demonstrate interactions passing continuously from Poissonian to super-Poissonian and up to thermal statistics, unveiling a surprising manifestation of Bohr's Correspondence Principle: the transition from quantum walk to classical random walk on the free-electron energy ladder. The electron walker serves as the probe in non-destructive quantum detection, measuring the photon-correlation $g^{(2)}(0)$ and higher-orders $g^{(n)}(0)$. Unlike conventional quantum-optical detectors, the electron can perform both quantum weak measurements and projective measurements by evolving into an entangled joint-state with the photons. Our findings suggest free-electron-based non-destructive quantum tomography of light, and constitute an important step towards combined attosecond-temporal and sub-Å-spatial resolution microscopy.

**Introduction**

Understanding light as a wave phenomenon was firmly established in the early days of the 19$^{th}$ century, forming the foundations of Maxwell's equations (*1*), the entirety of electrodynamics, and the study of light–matter interactions. Later discoveries unveiled the quantum-particle nature of light: starting from the photoelectric (*2*) to the discovery of entanglement many decades later (*3*), constituting modern quantum optics (*4,5*). Yet, entire areas of science still find the wave description of light fully sufficient. This is true for light in the optical range and for all other forms of electromagnetic waves. For example, classical electromagnetic waves determine the dynamics of charged particles in numerous applications, from free-electron lasers and synchrotrons to radars, communication satellites, and even microwave ovens. In fact, light remains purely a wave phenomenon also at the frontier of free-electron–light experiments, where laser-driven electron acceleration (specifically, wakefield accelerators (*6,7*) and dielectric laser accelerators (DLA) (*8-11*)) still consider the particle motion using a wave description of the laser. Thus far, the quantum-particle nature of light in its interactions with free electrons remained hidden, without any direct consequences. This is in sharp contrast to the quantum nature of the *electron*, which is now regularly observed in ultrafast transmission electron microscopy (*12-15*). The critical role of the electron quantum wavefunction in its interaction with light is already well understood (*16-21*): the electron exchanges integer multiples of the photon energy, as shown in the theoretical description of the technique called photon-induced nearfield electron microscopy (PINEM) (*12, 16-17*). However, even in this case – where the electron must be treated quantum-mechanically – the light still acts as a classical wave in all experiments to date.

Here we present the first experiment showing that quantum statistics of photons alters their interaction with free electrons, i.e., we demonstrate the first free-electron–light interaction wherein light cannot be described as a wave phenomenon. Moreover, we use the electron-light interaction to measure the photon statistics of the light. This capability spawns from a fundamental aspect of quantum optics: the interaction with light can create entanglement between the light and the interacting object. Consequently, the joint free-electron–light state becomes a non-separable state from which we can extract the photon statistics by measuring the electron energy spectrum. Our experiment demonstrates this concept on amplified light, exhibiting different photon statistics for varying amplification regimes. The free-electron interaction enables us to characterize the amplifier output, showing its transition from coherent photon state with Poissonian statistics in the regime of deep saturation to super-Poissonian statistics due to amplified spontaneous emission and eventually to a thermal state with Bose-Einstein statistics. Looking forward, our experiments offer a pathway for using free electrons for quantum state tomography of light, with high resolution in both time and frequency (*22*).

**Bohr's Correspondence Principle: from free-electron quantum walk to random walk**

Our experiment shows that the measured electron acts as a "walker" performing a generalized quantum/classical walk (Fig. 1) on the ladder of energy levels (separated by the single-photon energy $\hbar\omega$), similar to an upside-down (quantum) Galton board experiment (Fig. 1a). At each infinitesimal step, the electron "walker" either remains in its current energy level or moves to a lower (higher) energy level by emitting (absorbing) a photon. Each step either maintains a well-defined phase between the electron energy states, as in interactions with coherent light (Fig. 1 left panels), or washes out the phase and acts in a random fashion, as in interactions with thermal light (Fig. 1 right panels). We observe the walk dynamics by measuring the electron energy spectrum as a function of illumination power for different photon statistics (Fig. 1c,d). The measurement shows that when light acts as a wave (coherent state), the resulting electron dynamics is also that of a wave – a pure quantum walk. In contrast, when the particle-nature of light becomes dominant (super-Poissonian and thermal statistics), the resulting electron dynamics is also that of a particle – a classical random walk. This emergence of the classical dynamics of random walk from the interaction of a free electron with super-Poissonian light represents a new manifestation of the Correspondence Principle between classical and quantum physics, which has been one of the cornerstones of quantum mechanics since its earliest days (*23*).

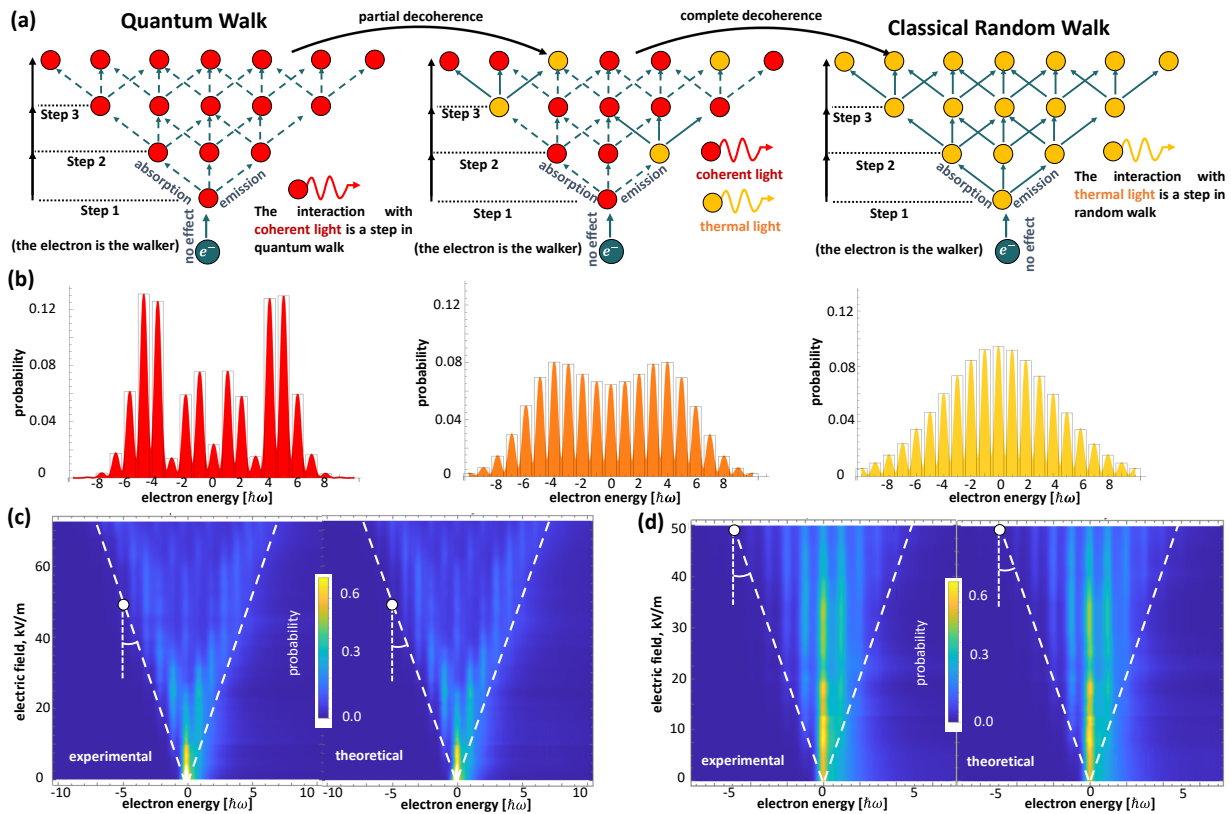

**Figure 1| Free-electron–light interactions imprint the quantum photon statistics on the electron energy spectra: demonstrating the transition from quantum walk to classical random walk of a free electron. (a)** The electron "walker" performs consecutive interactions with the photons. At each infinitesimal step, there are three possibilities: 1) the electron remains in its current energy level; 2) the electron emits a photon, moving to a lower energy level; 3) the electron absorbs a photon, moving to a higher energy level. Each such emission/absorption either maintains quantum coherence with a well-defined phase between the electron energy states, as in interactions with coherent light (red), or collapses and acts in a random fashion, as in interactions with thermal light (yellow). The collapse washes out the phase between the electron energy states. The transition from quantum walk to random walk is expressed by the continuous transition in the photon statistics from Poissonian (coherent state) to super-Poissonian and all the way to a thermal light state. **(b)** The electron walker theory exactly matches with the Q-PINEM theory, in which

the shape of the photon statistics determines the degree of decoherence of the electron. The resulting electron energy spectra show either quantum walk, random walk, or an intermediate case displaying partial decoherence, depending on the photon statistics (see also Fig. 3 for a measurement of the continuous transition). **(c,d)** Electron energy spectra for coherent and thermal states evolving with the electric field amplitude. Theory gives an excellent agreement with experimental results, measured in the setup presented in Fig. 2, with more data shown in the Supplementary Fig. S12. The white circles emphasize that we observe the same electron energy width for the same driving light intensity, regardless of the quantum statistics. This comparison reveals an important finding: the electron interactions with thermal states are as efficient as interactions with coherent states. This fact shows that the interaction of super-Poissonian light with a free electron does not cause the electron to decohere in time or space. Rather, the electron remains in a coherent superposition of energy states that become more entangled with the light for stronger super-Poissonian statistics, with maximum entanglement for thermal light.

We begin by formulating the theory of light–free-electron interaction and its dependence on photon statistics. The theory, henceforth called the electron walker theory (explained in Fig. 1a,b), captures the entire range of experimental parameters, including both limiting cases of coherent and thermal light: For the interaction with coherent-state light (Fig. 1 left panels), the electron undergoes pure quantum walk in energy-space (as shown in Ref. *13*), and for the interaction with thermal light (Fig. 1 right panels), the electron undergoes pure random walk. The intermediate cases of interactions with super-Poissonian light all correspond to mixtures of quantum walk and random walk, showing a continuous transition between the two phenomena. We show in Supplementary Materials (SM) S1 that despite its simplicity, the walker theory provides the same results as the quantum-optical generalization of PINEM (Q-PINEM), proposed theoretically in Refs. *24-25* (in particular, Fig. 1d precisely matches the prediction in Ref. *25*). The comparisons in Fig. 1c,d show that the theories successfully describe the measured electron energy spectra, the experimental details of which we describe below.

The agreement of both walker and Q-PINEM theories with the experimental data in Fig.1c,d allows us to draw decisive conclusions about the role of quantum decoherence in the interaction.

The photon statistics directly determine the degree of decoherence of the electron walker: broader photon distributions (e.g., super-Poissonian) increase the free-electron decoherence in energy (*26*) during the electron walk. The limit of thermal light can be understood exactly as the complete decoherence of the electron quantum state at every step of the walk – showing the emergence of classical random walk from the decoherence of quantum walk. The phenomenon of quantum walk has been observed experimentally in a wide range of physical systems (*27-32*). We note an intriguing fundamental difference: the quantum-to-classical transition of the electron walker we observe arises from the photon distribution rather than from dephasing and disorder-induced decoherence as in many other systems (*29, 32-36*).

**Experimental setup: high-efficiency silicon-photonic electron–light coupler**

Our experiment achieves an efficient electron–light interaction via two critical ingredients: photonic cavities (*37-38*) and phase-matching between the electron wavefunction and the light wave (*39*) inspired by phase-matching of classical electron–light interactions (*40*). Here, we employ quasi-phase-matching in custom-made silicon-photonic nanostructures to create efficient free-electron–quantum-light interactions inside a transmission electron microscope (TEM) (Fig. 2a). Our nanostructures are inspired by miniaturized electron accelerator (*8-11*). We utilize photonic inverse design methods (SM S8.3) to design and optimize a high-efficiency electron–light coupler (Fig. 2d) (*11,41*), operating at a wavelength of 1064 nm and electron kinetic energy of 189 keV (Fig. 2). We modify the quantum statistics of the output light by controlling the input power to a fiber amplifier and then couple the light into the nanostructure inside the TEM using two cylindrical lenses (SM S8.1).

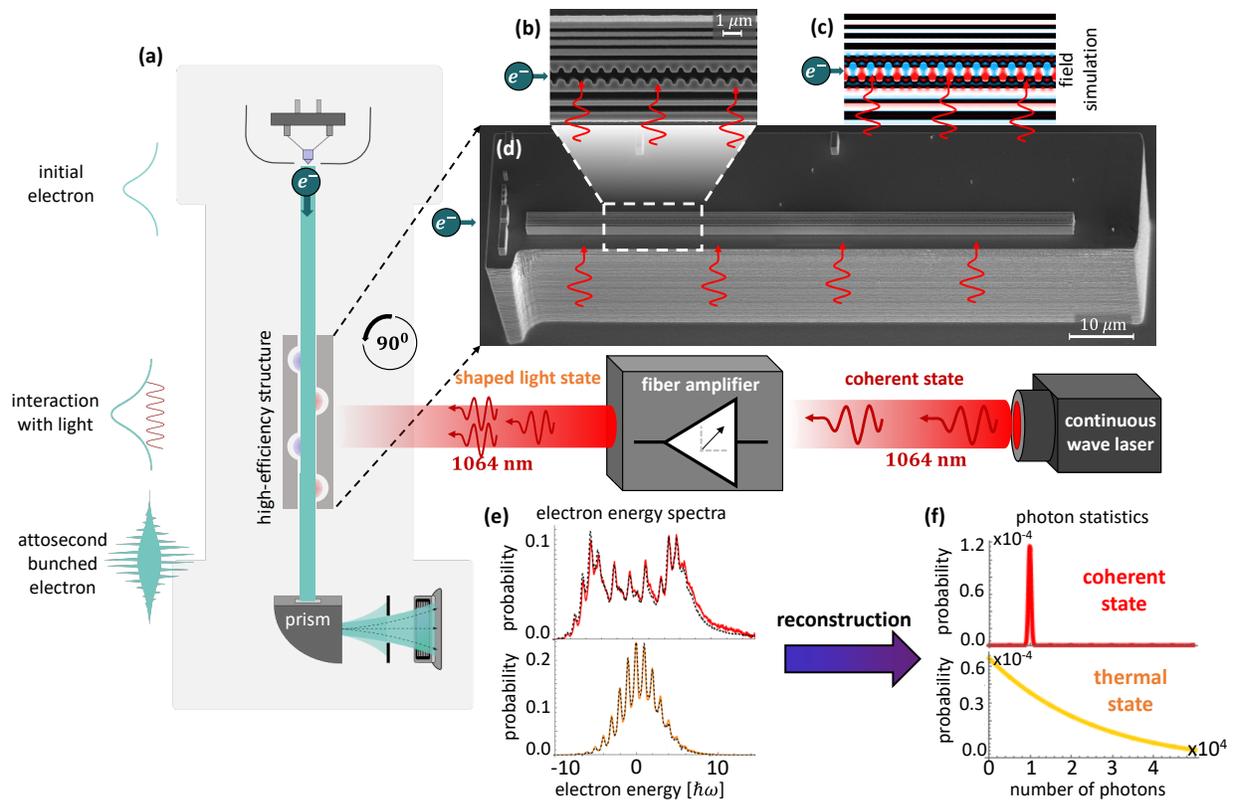

**Figure 2 | Free-electron interaction with light of different photon statistics in a high-efficiency silicon-photonic nanostructure. (a)** We use a high-efficiency electron–light coupler in a transmission electron microscope (TEM) to facilitate efficient interaction of free electrons with continuous-wave (CW) light. With this, we observe the effect of photon statistics in electron–light interactions. The electron energy spectrum is measured with an energy resolution better than the single-photon energy, using electron energy loss spectroscopy (EELS). We vary the photon statistics continuously with the help of the fiber amplifier from Poissonian to super-Poissonian and up to thermal. **(b)** Scanning electron microscope image of the coupling structure: light and electrons are efficiently coupled using a resonating quasi-phase-matched structure consisting of a periodic channel and a Bragg mirror. **(c)** Simulation of the longitudinal quasi-phase-matched electric field in the nanostructure, optimized with photonic inverse design methods for efficient free-electron–light interaction. **(d)** Image of the entire nanostructure in 3D; more info in SM S8.2 and Supplementary Fig. S17. **(e)** Electron energy spectra for interactions with coherent and thermal light; the colored curves denote the theory (fit explained in Supplementary Fig. S12 and SM S4) and dashed black curves denote the experiment, showing almost perfect overlap. **(f)** The measured electron spectra are used to extract the photon statistics shown here. The slight asymmetry between gain and loss observed in panel (e) and in Fig. 3d is explained in SM S4.3.

Our experiment realizes strong electron–light interactions, i.e., each electron exchanges multiple photons with the light field. Such strong interactions were previously only realized with intense laser pulses synchronized with photo-emitted free electrons (*8-14, 37-39, 42*). In contrast with such intense laser pulses that can be considered as coherent states (classical waves), all other states of light are usually not so intense. The low intensity poses a challenge for investigating their interactions with free electrons. We thus developed nanostructures for enhancing the interaction efficiency, reaching the desired strong electron–light interaction with lower intensity light, which is especially important for quantum light. The interaction efficiency is high enough to enable the use of continuous-wave (CW) light, while still maintaining the strong interaction. Recent experimental works have shown that even weak CW electron–light interactions (up to one photon absorbed/emitted by the electron) have intriguing applications (*43-46*). Our experiment offers an avenue for taking these ideas forward to regimes of stronger interactions with CW light, and specifically enables us to probe the quantum statistics of the photons.

**Free-electron–quantum-light interactions**

In all free-electron experiments to date, the electron dynamics has been accurately captured by its interaction with classical electromagnetic fields. The dynamics of a single-electron wavefunction has been fully described by the time-dependent Schrödinger equation (or the Dirac equation in the more general relativistic case) with the classical vector **A** and scalar $V$ potentials:

$$i\hbar \partial_t |\psi\rangle_\text{el} = \left[(\mathbf{p} + e\mathbf{A}(t))^2/2m + eV(t)\right]|\psi\rangle_\text{el}, \qquad (1)$$

where $\mathbf{p} = -i\hbar\nabla$ is the momentum operator, $e$ and $m$ are the electron charge and mass, and $|\psi\rangle_\text{el}$ denotes the electron quantum state. This description was justified so far because in the quantum picture, classical electromagnetic fields can be described as coherent states $|\alpha\rangle_\text{ph}$ (*47*), and intense coherent states stay approximately unchanged under interactions. Consequently, in a full quantum description, the joint electron–photon state $|\Psi\rangle_\text{el–ph} \approx |\psi\rangle_\text{el} \otimes |\alpha\rangle_\text{ph}$ remains *separable* after the interaction with an intense classical field.

In contrast, for some of the photonic states we consider here, the joint electron–photon state after the interaction is *non-separable*, i.e., entangled (*48*). This situation cannot be described by the commonly used theoretical analysis of a time-dependent Schrödinger equation with non-quantized potentials (Eq. (1)) but requires a quantum-optics theory. Below, we present the formulation of the quantum interaction of light with a highly-paraxial free-electron beam, i.e., the Q-PINEM theory (*22, 24-25*), which has the scattering matrix:

$$S = \exp(g_\mathrm{q} b a^\dagger - g_\mathrm{q}^* b^\dagger a). \tag{2}$$

$g_\mathrm{q}$ is the quantum coupling constant; $a, a^\dagger$ are the photonic annihilation and creation operators; and $b, b^\dagger$ are the free-electron energy ladder operators describing the electron losing or gaining a single-photon energy quantum. The resulting joint electron–photon state is generally *non-separable*, and its density matrix can be written as follows:

$$\rho_\mathrm{el-ph} = \sum_{n,m=0}^{\infty} \rho_\mathrm{ph}(n,m) |\Psi^n\rangle_\mathrm{el-ph} \langle\Psi^m|_\mathrm{el-ph}, \tag{3}$$

where $\rho_\mathrm{ph}(n,m) = \langle n|\rho_\mathrm{ph}|m\rangle$ is the density matrix element of the light in Fock space. Each pure state component $|\Psi^n\rangle_\mathrm{el-ph}$ is a superposition of electron–photon product states. The experimental conditions allow us to consider the initial electron as a single-energy state $|E_0\rangle_\mathrm{el}$ with energy uncertainty smaller than the energy $\hbar\omega$ of a single photon of the field with which it interacts. In this case, we can express the joint electron–photon state as:

$$|\Psi^n\rangle_\mathrm{el-ph} = \sum_{k=-\infty}^{\infty} c_k^n |E_0 - k\hbar\omega\rangle_\mathrm{el} \otimes |n+k\rangle_\mathrm{ph}, \tag{4}$$

where $|n+k\rangle_{\text{ph}}$ is a Fock state of light with $n+k$ photons, and $|E_0 - k\hbar\omega\rangle_{\text{el}}$ is the electron state with energy $E_0 - k\hbar\omega$. Eq. (4) shows that the free-electron–light interaction can create non-separable quantum states. By solving Eq. (2) for the state evolution, we find the coefficients $c_k^n = |g_q|^{-1}(-1)^n e^{i\varphi_g k} W_{n+\frac{1}{2}(k+1),\frac{1}{2}|k|}(|g_q|^2)/\sqrt{(n+k)!\,n!}$, where $W$ is the Whittaker function (SM S2). In the limit of weak interaction and a large number of photons, the coefficients become $c_k^n \approx e^{i\varphi_g k} J_k(2|g_q|\sqrt{n})$ (22, 25), reminiscent of the PINEM theory (16-17).

**Extracting the photon statistics from the electron energy spectrum**

The photon statistics encapsulated in the photonic density matrix $\rho_{\text{ph}}$ can dramatically change the measured electron energy spectrum (22, 24-25). We utilize this concept to experimentally demonstrate how the statistics of light interacting with free electrons is imprinted on the corresponding electron energy spectra. The probability $P_k$ to measure an electron energy shifted by $k$ photons depends on $\rho_{\text{ph}}$ through $P_k = \sum_n |c_{-k}^{n+k}|^2 \rho_{\text{ph}}(n+k, n+k)$. By inverting this relation, we can extract the photon statistics from the measured electron energy spectrum as shown in Fig. 3. Specifically, Figs. 3a,b show five electron energy spectra (experiment and theory, respectively) throughout the continuous transition between amplified spontaneous emission and amplified coherent light; Fig. 3c shows the extracted photon statistics for each case.

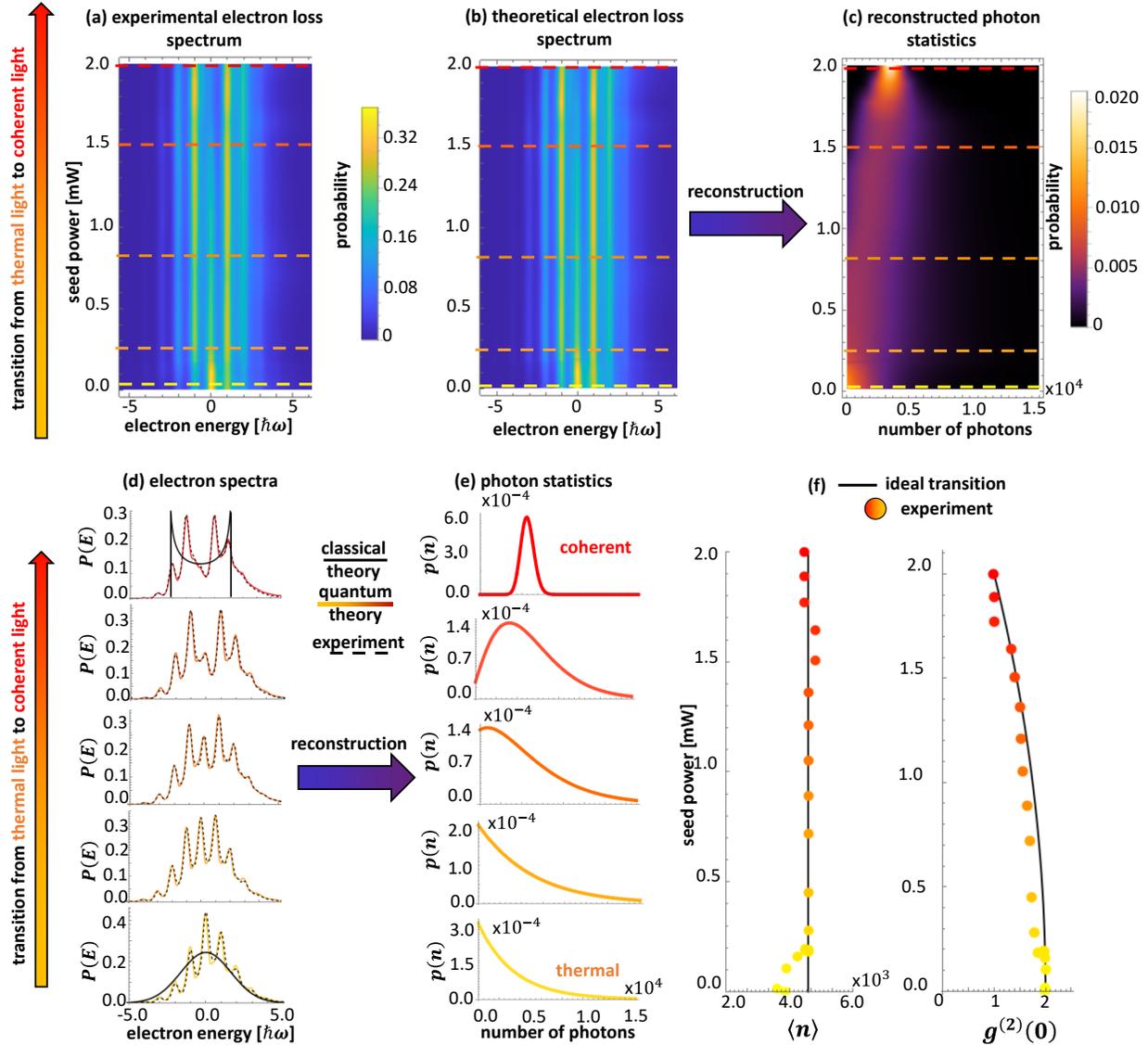

**Figure 3| Experimental reconstruction of photon statistics from electron energy spectra. (a)** The measured electron energy spectra for a range of fiber amplifier parameters, showing a continuous transition from coherent Poissonian light in the regime of deep saturation to super-Poissonian and eventually thermal light, created by amplified spontaneous emission. The average photon number is the same in all cases. **(b)** The corresponding theoretical calculation of the electron spectra is based on the Q-PINEM theory (Eqs. (2-4)) with the quantum-optical amplifier model (Eq. (5), SM S3). **(c)** We reconstruct the photon statistics of light from the experimentally obtained electron spectra (a) by using the fitted parameters of the theoretical model (b) and substitute them in the amplifier model (SM S4). **(d)** Selected cases of measured electron energy spectra following the interaction with different states of light. The data (dashed black curve) show an excellent match to the Q-PINEM theory (colored solid curve). By contrast, simulations using a point electron (solid black curve) do not at all agree with the quantum predictions, indicating that the effects cannot be obtained from an incoherent point electron mixture/collapse in time (SM S7). **(e)** Reconstructed photon statistics corresponding to the selected cases of (d). **(f)** The average number of photons $\langle n \rangle$ and second-order correlation (degree of coherence) $g^{(2)}(0)$ extracted from the measurements. The black curves correspond to the theory (SM S2) with the same parameters.

To unveil the role of photon statistics, we gradually vary the state of light (Fig. 3a) from coherent to thermal and reconstruct the statistics at each stage (Fig. 3c). We change the statistics by varying the power of a coherent laser beam seeding the fiber amplifier, thereby varying the gain saturation and output power (*49-50*). We keep the optical power entering the electron microscope constant by using an attenuator. This way, we change the photon statistics without changing the average number of photons, as shown in Fig. 3f. Therefore, the measurement in Fig. 3a directly corresponds to the quantum-to-classical transition of the free-electron quantum/classical walk shown in Fig. 1a,b. As explained in SM S4.4 and Supplementary Fig. S15, our observations cannot arise from the lack of spatial/temporal optical coherence nor from spatial or temporal inhomogeneity of the light driving the interaction.

We emphasize that the limiting case of a coherent state corresponds to an approximately separable $\rho_{el-ph}$, while the other limiting case of a thermal state corresponds to a non-separable $\rho_{el-ph}$. This can be directly seen in Eq. (3), where $\rho_{ph}$ of thermal light is diagonal, making the joint $\rho_{el-ph}$ non-separable. Both limiting cases, and all cases in between, can be accurately captured by the Q-PINEM theory (Eqs. (2-4)). To quantify the quantum-optical state of light, we also extract the corresponding second-order degree of coherence $g^{(2)}(0)$ from each electron spectrum (Fig. 3f) as first proposed in Ref. (*25*). Moreover, since the entire photon distribution can be extracted (Fig. 3c), it directly provides all the photon-number moments (*22*) $\langle n^m \rangle$ and thus also the higher-order $g^{(n)}(0)$ (*25*).

The ability to control and measure the photon statistics during light–free-electron interaction offers a variety of applications. As a proof-of-concept application, we characterize the amplifier output when it undergoes a transition from the limit of Poissonian statistics in the regime of deep saturation to super-Poissonian statistics created by amplified spontaneous emission (*5*). The amplifier has a coherent input seed of amplitude $\alpha$ and a saturable gain $\mathcal{G} = \mathcal{G}(|\alpha|^2)$, generating light with the following photon statistics (SM S3):

$$\rho_{\rm ph}(n,n) = e^{-|\alpha|^2} \frac{1}{\mathcal{G}} \left(1 - \frac{1}{\mathcal{G}}\right)^n L_n\left(-\frac{|\alpha|^2}{\mathcal{G}-1}\right), \qquad (5)$$

where $L_n(x)$ is the $n$-th Laguerre polynomial. By tuning $\alpha$ and $\mathcal{G}$, the photon statistics of the output light can be continuously tuned between amplified coherent state (for $|\alpha| \gg 1$, saturated $\mathcal{G}$, approaching a Poissonian photon distribution) to thermal amplified spontaneous emission (for $|\alpha| \approx 0$, linear $\mathcal{G} \gg 1$, approaching a thermal photon distribution).

We extract the amplifier gain curve from the analysis of the electron energy spectrum (Supplementary Fig. S9). We also perform two additional independent measurements of the amplifier gain that both show good agreement (Supplementary Fig. S9e): from the optical spectra and from direct power measurements. This comparison supports our estimation of the photon statistics from the amplifier model. Furthermore, we can use the gain curves to estimate the quantum coupling constant $g_q$ up to the coupling efficiency of the light to the structure (elaborated in SM S4). In this fashion, the free-electron quantum-optical detection can be applied to extract the ultrafast quantum statistics emerging from amplifiers operating on sub-ps and even sub-fs timescales.

**The free electron as a probe of quantum weak measurement**

The free electron can serve as the probe in quantum non-destructive detection of the state of light $\rho_{\rm ph}$ (51), as suggested theoretically in the context of PINEM (52). The electron plays the role of the measurement device/pointer and the light plays the role of the system to be measured. Specifically, the electron detection in the electron energy loss spectrometer (EELS) is a strong projective measurement of the electron that can be modeled by a set of projection operators $M_k = $

$|E_0 - k\hbar\omega\rangle\langle E_0 - k\hbar\omega|$. Following a projection by $M_k$, the state of light changes from $\rho_{\text{ph}}$ to a new state denoted by $\rho_{\text{ph}}^k$ (the superscript $k$ indicates the correlation with the electron measurement). The density matrix of this light $\rho_{\text{ph}}^k$ may or may not differ significantly from the original state $\rho_{\text{ph}}$, depending on the level of entanglement with the electron. The difference is quantified by the fidelity (53) $F_k = \left[\text{tr}\left\{\sqrt{\sqrt{\rho_{\text{ph}}}\rho_{\text{ph}}^k\sqrt{\rho_{\text{ph}}}}\right\}\right]^2$ between $\rho_{\text{ph}}$ and $\rho_{\text{ph}}^k$ for each value of $k$ (plotted in Fig. 4a). High fidelity ($F_k \to 1$) means that the light is mostly unchanged by the measurement, whereas low fidelity ($F_k \to 0$) means that the light is substantially changed by the measurement.

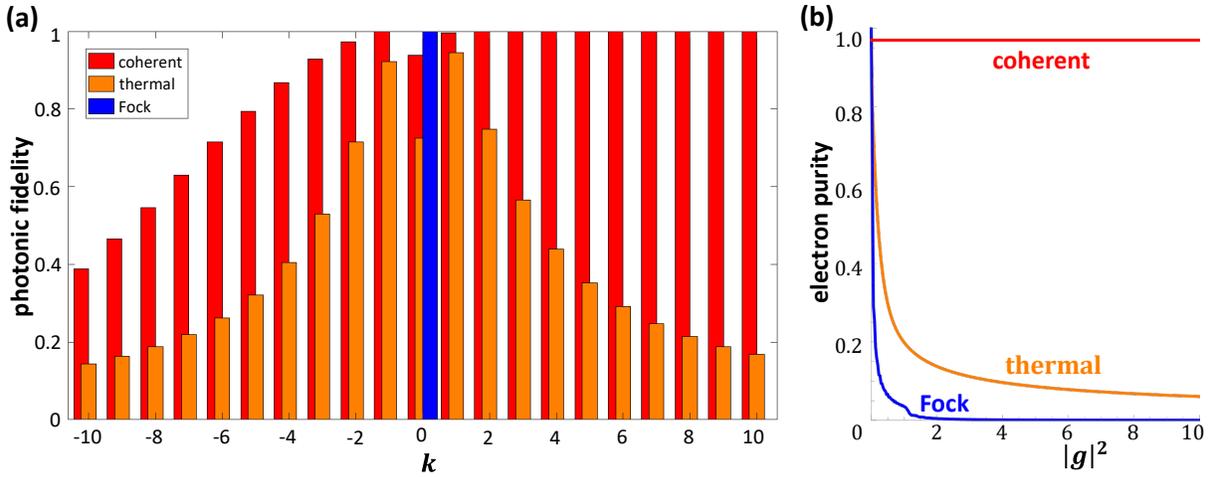

**Figure 4| Photonic fidelity and electron purity following a free-electron–quantum-light interaction. (a)** Calculated fidelities between the photonic state before and after its interaction with a free electron, which subsequently undergoes a projective measurement. The fidelity is plotted as a function of the energy state $|E_0 - \hbar\omega k\rangle$ in which the electron is detected. We present the fidelity for different states of light, all with an average photon number $\langle n \rangle = 100$ and quantum coupling $g_q = 0.1$. The fidelity is high for a coherent state, indicating that for small-enough $g_q$, the electron can serve as a probe of quantum weak measurement. For the thermal state, the fidelity is much lower, indicating a behavior closer to strong projective measurement. As a comparison, we also present the fidelity for a Fock state, for which the measurement is projective (fidelity of zero) for any $k \neq 0$. The fidelity asymmetry for the coherent case is explained by the fundamental difference between subtracting or adding photons to a coherent state (SM S2). **(b)** Electron purity following its interaction with different states of light as a function of the classical coupling strength $|g|^2 = |g_q^2|\langle n\rangle$. The electron stays pure after an interaction with a coherent state, but its purity diminishes much faster for a thermal state, as an indication of emerging entanglement between the electron and light.

The electron can act as a probe for weak quantum measurements (*54-58*) with a near-unity fidelity for interactions with coherent light. i.e., the electron performs a measurement of the light, yet does not substantially affect the quantum state of the light, which is why the process is a weak quantum measurement. In contrast, the interaction with thermal light causes a substantial reduction in the fidelity, implying that the electron probe acts more like a projective measurement that does alter the state of light. Therefore, the same electron probe manifests both extremes: weak measurement and projective measurement (Fig. 4a), depending on the state of the measured system. To test this observation, a direct measurement of the outgoing state of light after the interaction could be performed by using high efficiency coupling into/out of the nanophotonic structure (SM S5.4), which is beyond the scope of this work.

To show theoretically the role of the electron probe in the measurement process, we calculate the purity of the post-interaction electron (Fig. 4b), which shows that the electron maintains a purity of unity for coherent light, but the purity substantially decreases for thermal light. Such a substantial reduction in purity can be understood as the collapse of the electron wavefunction in the energy domain. The walker theory we developed (SM S1) shows that this description accurately captures the measured effect and precisely matches the Q-PINEM theory. Through the walker theory, the electron–light interaction can be understood as consecutive infinitesimal steps at which the electron quantum state partially collapses in energy, with full (no) collapse occurring for thermal (coherent) light, resulting in a pure random (quantum) walk. We emphasize that the electron wavefunction never collapses in time and space but only in the energy domain. Spatially and temporally, the electron remains a coherent wavefunction that extends over many cycles of the field, as evident by the emergence of discretized energy peaks (e.g., in Fig. 3d and in all other figures), occurring even with thermal light. A collapse in time or space, if it ever appears, would manifest as an incoherent mixture of point particles interacting with the light field, which cannot result in discrete energy peaks (Fig. 3d solid black curves, SM S7). In contrast, the collapse in energy occurring in our experiment causes the electron to split into multiple extended wavefunctions, in a joint quantum state entangled with the state of light.

**The role of quasi-phase-matching and connection to the Smith-Purcell effect**

The key to the efficient electron–light coupling that enables our entire experiment is matching the electron velocity with the phase velocity of light trapped in the nanostructure. This phenomenon is known as *quasi-phase-matching between free electrons and light* (SM S5), also known as the inverse Smith–Purcell effect (*9,59*). Our new realization of this effect goes beyond conventional realizations of inverse Smith–Purcell experiments in three ways: 1.) using CW light rather than laser pulses, 2.) having an electron wavefunction rather than a classical point electron, as proposed theoretically in Refs. (*60-62*), and 3.) varying the photon statistics so that the experiment deviates from the semi-classical theory treated in Refs. *60* and *61*, and generally cannot be described by any time-dependent Schrödinger equation (Eq. (1)). To highlight and quantify the quasi-phase-matching, we scan over the electron energy (Fig. 5), searching for the optimal electron–light coupling conditions. We find the optimal acceleration voltage to be 189 keV and estimate the effective interaction length to be 56 μm, shorter than the structure length of 84 μm. This deviation arises from the 51 μm FWHM transverse intensity profile of the light beam (SM S8.1).

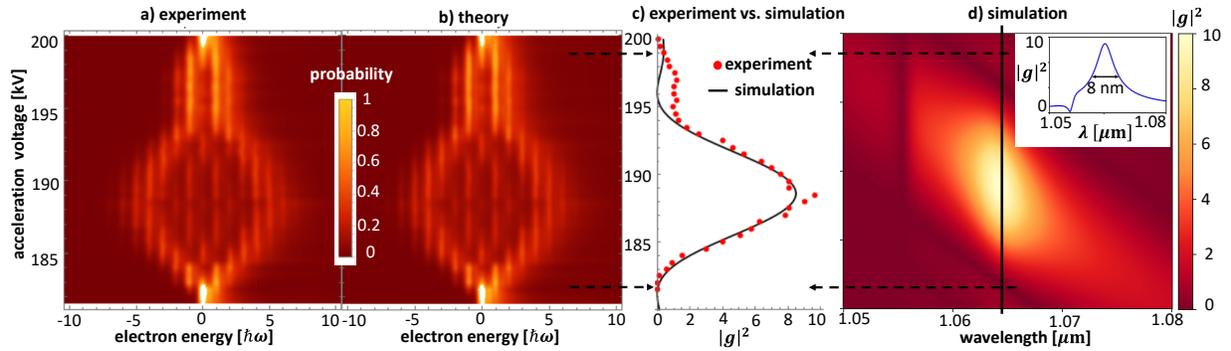

**Figure 5| Measurement of the quasi-phase-matching condition. (a)** Electron energy spectra measured for a range of electron kinetic energies, **(b)** showing a good agreement with theory. **(c)** We extract the classical interaction strength $|g|^2 = |g_q|^2 \langle n \rangle$ and effective interaction length from the data by comparing it with the quasi-phase-matching theory (SM S5). **(d)** Simulation of $|g|^2$ as a function of electron kinetic energy and laser wavelength, showing the resonant nature of the quasi-phase-matched interaction. Supplementary Fig. S14 presents a version of panel (d) for a longer structure, allowing a better frequency resolution for isolating a single mode in the structure.

**Discussion and outlook**

The underlying quantum-optical theory describing our experiment is best understood in light of recent theoretical advances (*24-25, 62*), that have caused a paradigm shift in how we think about the interaction of free electrons and light. Specifically, these and other works show how the interaction becomes sensitive to the photon statistics (*22, 24-25, 62-65*) and provide new opportunities for shaping electron wavefunctions(*63*). These promising ideas open novel avenues of research in free-electron quantum optics, suggesting new concepts such as electron–light entanglement that can induce electron–electron entanglement (*24, 66-67*). This entanglement can in turn be utilized to create new quantum light sources (*64*), improve cathodoluminescence techniques (*66, 68*), and develop future Cherenkov detectors for particle physics (*69*). Since all previous free-electron–light experiments worked with classical (coherent state) light, our experiment is a proof-of-concept for the predictions on interactions between free electrons and non-classical light, thereby paving the path for these exciting applications.

On the technical side, our experiment highlights new applications of silicon-photonic nanostructures in the field of quantum optics, in which free electrons provide a novel mechanism to extract the quantum photon statistics. This concept can be extended into full quantum state tomography of light (*22*) by using Ramsey-type experiments (*18*). Intriguingly, this approach to quantum state tomography does not have to destroy/absorb the measured light. Such free-electron-based quantum-optical detectors can have extremely broad bandwidths and the ability to selectively detect individual modes by using phase-matching (SM S5). The outlook of such quantum-optical detection techniques depends on the efficient coupling of light into the nanostructure that performs the electron–light interaction; the necessary coupling capabilities are already available as on-chip technology (*11, 70*).

Before closing, we note that our small electron probe (e.g., 30 nm diameter, SM S8.1) is sensitive to minute transverse changes in the field across the channel of the nanostructure, mapping the field at deep subwavelength resolution. Thus, our methods can help identify optimal operation conditions of these nanostructures for different applications, such as laser-driven electron acceleration. Future experiments would enable comparing performances of different electron accelerators: between the more traditional pillars-based nanostructures (*71*) and ones based on inverse design optimization (*11*). We can characterize the spectral response with a resolution limited only by the excitation linewidth, being extremely narrow if using a tunable CW source.

Last but not least, over the last decade, coherent shaping of spatial electron wavefunctions (*42, 72-76*) has opened new avenues in electron microscopy such as electron magnetic circular dichroism (*73*) and aberration corrections (*77*). More recently, coherent temporal modulation of electron wavefunctions (*13, 18, 46, 63, 78-79*) has excited new ideas for ultrafast light-matter interactions and free-electron coherent control (*18, 19, 80*). Especially interesting are the applications of modulation by CW light, showing better phase contrast (*43*), PINEM by electron post-selection (*44*), plasmon excitation mapping (*45*), and advances toward attosecond-resolution metrology (*46*). However, reaching *strong* electron modulation in a CW operation has been a long-standing challenge that remained unanswered. Our experimental demonstration paves the way towards equipping state-of-the-art microscopes with CW silicon-photonic light couplers. The modulation capabilities may be further extended to combined spatial and temporal shaping, by structured light–electron nanophotonic coupler and/or specially shaped light beams (*81*). Such dream microscopes could operate at simultaneous sub-Å-spatial and few-attosecond-temporal resolution (*18-20, 82*) and lead to breakthrough experiments in some of the hardest fundamental problems: providing direct observation of molecular excitation dynamics, electron transport phenomena, transient sub-cycle phenomena in optical nonlinearities, ultrafast plasma oscillations, and many more effects that cannot be explored by other means.

*A related work that also shows continuous-wave PINEM using silicon photonic structures appeared in parallel with our work (83) .*

# Imprinting the quantum statistics of photons on free electrons

Raphael Dahan[†], Alexey Gorlach[†], Urs Haeusler[†], Aviv Karnieli[†], Ori Eyal, Peyman Yousefi, Mordechai Segev, Ady Arie, Gadi Eisenstein, Peter Hommelhoff, and Ido Kaminer

# Supplementary Material

## Contents

# S1. The theory of quantum and random walk

In this section, we consider the interaction between light and free electrons and show how it can be described in terms of quantum walk, random walk, and the combination of the two.

## S1.1. The connection between the interaction of a free electron with coherent-state light to the quantum walk model

Let us consider the following model describing the interaction between light and an electron in a discrete energy ladder. The energy ladder forms a synthetic dimension in which the electron makes discrete steps. At each step, the electron has the following three options: 1) absorb a photon and climb up one step in the energy ladder; 2) emit a photon and go down one step in the energy ladder; 3) no interaction, retaining its previous energy. Moreover, in this simple model, we consider that the emission and absorption have the same probabilities (which is a good approximation for large photon numbers). We further assume that the light is unperturbed by the interaction so we can consider the probabilities as constants throughout the walk. The electron wavefunction can be written as a column vector of complex number amplitudes, where each component of the column represents a monoenergetic state:

$$|\psi\rangle = (\ldots, \psi_{-1}, \psi_0, \psi_1, \ldots)^T, \quad (S1.1)$$

which denotes $|\psi\rangle \equiv \sum_k \psi_k |E_0 + k\hbar\omega\rangle$, where $|E_0 + k\hbar\omega\rangle$ is a monoenergetic state with the energy $E_0 + k\hbar\omega$ ($E_0$ is the initial energy). After $n + 1$ steps, the $k^{\text{th}}$ component has the following form:

$$\psi_k^{(n+1)} = \sqrt{1 - 2|c|^2}\psi_k^{(n)} + c\psi_{k+1}^{(n)} + c^*\psi_{k-1}^{(n)}, \quad (S1.2)$$

where $c$ is a complex number describing the transition amplitude of the interaction between light and the free electron. The recurrence relation Eq. (S1.2), together with the initial condition $\psi_k^{(0)} = \delta_{k0}$, gives the following probability to absorb (or emit) $k$ photons after $N$ steps:

$$P_k = \left|\psi_k^{(N)}\right|^2 = |J_k(2|\beta|)|^2, \quad (S1.3)$$

where $J_k$ is the Bessel function of the first kind of order $k$, and $|\beta| = |c|N$. This result proves the equivalence between the quantum walk theory and free-electron interaction with coherent-state light because Eq. (S1.3), obtained from the discrete step quantum walk, precisely matches Eq. (S2.27), obtained from the quantum theory of photo-induced nearfield electron microscopy (Q-PINEM) in Section S2 and Section S4 and in Refs. S1, S2.

## S1.2 The connection between the interaction of a free electron with thermal-state light to the random walk model

To model the electron interaction with thermal light, we adopt a similar approach to the previous section. However, instead complex number amplitudes, we consider classical probabilities. As before, a free electron can interact at each step with the thermal light in the following ways: 1) absorb a photon; 2) emit a photon; 3) no interaction. We consider the state of the electron to be described by a classical vector of probabilities:

$$P = (\ldots, P_{-1}, P_0, P_1, \ldots)^T, \quad (S1.4)$$

where $P_k$ denotes the electron probability to be found in energy level $E_0 + k\hbar\omega$. Thus, Eq. (S1.4) is an analog of Eq. (S1.1) for random walk. After $n + 1$ steps, the electron probabilities are described by:

$$P_k^{(n+1)} = (1 - 2p)P_k^{(n)} + pP_{k+1}^{(n)} + pP_{k-1}^{(n)}, \quad (S1.5)$$

where $p$ is the probability to emit a photon. Eq. (S1.5) has the following solution if the initial conditions are $P_k^{(0)} = \delta_{k0}$:

$$P_k^{(N)} = e^{-2|\beta|^2} I_{|k|}(2|\beta|^2), \quad (S1.6)$$

where $|\beta|^2 = p \cdot N$. This result proves the equivalence between the random walk theory and free-electron interaction with thermal-state light because Eq. (S1.6), obtained from the discrete step random walk, precisely matches Eq. (S2.29), obtained from the Q-PINEM theory[S1-S3].

### S1.3 The continuous transition from quantum to random walk

Sections S1.1 and S1.2 described the limits of random and quantum walk as models for the interaction of free electrons with light in thermal and coherent states, respectively. In this section, we consider the continuous transition between these two limiting cases.

To describe the continuous transition, we employ the following model. The light is assumed to be in a mixture of coherent and thermal states, assuming $n_c$ photons from the coherent field, and $n_{th}$ photons from the thermal field. We find that the order of interactions with the free electron does not alter the final electron energy spectrum. Thus, we calculate the interaction by first applying the coherent-state photons. After the interaction with $n_c$ coherent photons according to Eq. (S1.3), the electron probability distribution of absorbing (or emitting) $k$ photons is:

$$P_k = \left|J_k\left(2\frac{|\beta|}{N}n_c\right)\right|^2, \qquad (S1.7)$$

This electron now interacts with $n_{th}$ thermal photons according to Eq. (S1.5). The final electron energy spectrum is calculated numerically using Eq. (S1.7) as the initial condition.

Fig. S1 shows that the resulting electron energy spectrum precisely matches the full Q-PINEM calculation using the state of light of the amplifier theory Eq. (S4.1). Matching the theories involves defining a thermality parameter $r_{th} \in [0,1)$, which we show to be connected to the amplifier gain and seed in Eq. (S4.2). The thermality $r_{th}$ is defined in the following way:

$$\begin{cases} n_c = \sqrt{1-r_{th}}N \\ n_{th} = r_{th}N \end{cases}, \qquad (S1.8)$$

with $r_{th} = 1$ giving pure thermal light, and $r_{th} = 0$ giving pure coherent light.

To summarize this section, we showed a perfect match between free-electron interaction with light to the theory of quantum-to-random walk transition. Therefore, the theory of this section provides a simpler description for electron–light interactions in place of the rigorous Q-PINEM theory (Section S2), applied on the amplifier light (Section S3), and combined in Section S4.1.

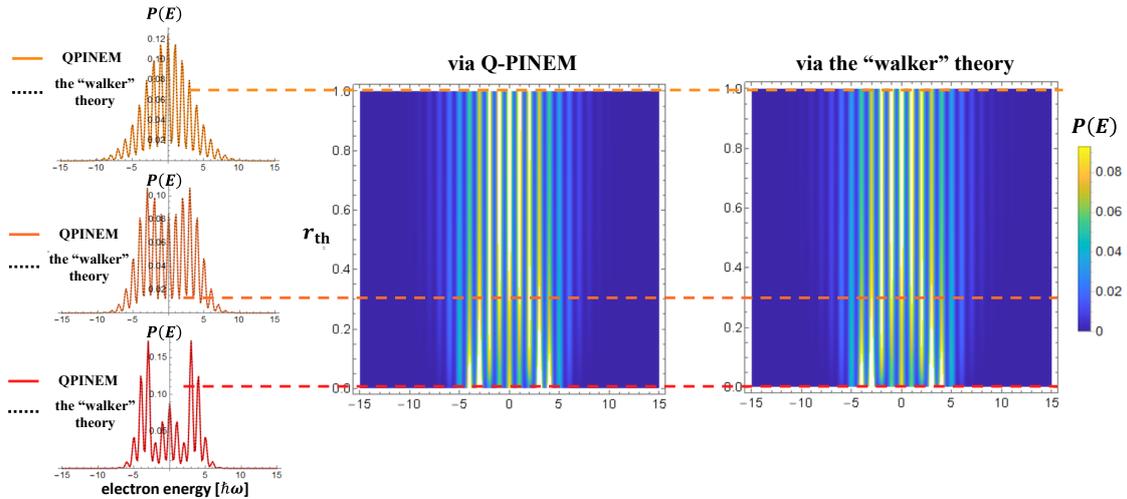

**Figure S1: Comparison between quantum-to-random "walker": theory and Q-PINEM theory, showing the equivalence of the two models**. The electron energy spectra as a function of the thermality coefficient $r_{th}$, calculated for both theories and showing the exact match between the results predicted by these different theories.

## S2. Interactions of free electrons with quantum light: the Q-PINEM theory

Here we discuss the quantum theory of the interaction between free electrons and quantum light (i.e., the theory of Q-PINEM). We discuss the implications of this general theory, such as electron–light entanglement, purity of the electron wave function, and the resulting electron energy loss spectra.

### S2.1. A general theory of the Q-PINEM interaction

In this section, we formalize the free-electron-light interaction. According to Refs. S1-S3, the scattering matrix of the interaction between free electrons and quantum light is described by:

$$S = \exp(g_q b a^\dagger - g_q^* b^\dagger a), \tag{S2.1}$$

where $g_q$ is the quantum coupling constant, $b \equiv e^{-i\omega z/v}$, $b^\dagger \equiv e^{i\omega z/v}$ are the electron energy ladder operators, and $a, a^\dagger$ are the photonic annihilation and creation operators. The final state after the interaction can be written as:

$$\rho_{\text{tot}}^{(f)} = S \rho_{\text{tot}}^{(i)} S^\dagger, \tag{S2.2}$$

where $\rho_{\text{tot}}^{(i)} = |E_0\rangle\langle E_0| \otimes \rho_{\text{ph}}$. We define $\rho_{\text{ph}}$ as the initial density matrix of the quantum light and $|E_0\rangle$ as the initial electron state with a well-defined energy $E_0$ (later, we will generalize it to an arbitrary initial electron state $\rho_{\text{el}}^{(i)}$).

### S2.2 The total density matrix following a Q-PINEM interaction

In this section, we calculate the joint density matrix of the electron and light $\rho_{\text{tot}}^{(f)}$ and show the electron-light entanglement. We rewrite the scattering matrix Eq. (S2.1) as a Taylor series:

$$S = e^{\frac{|g_q|^2}{2}} \sum_{m,l=0}^{\infty} \frac{(-g_q^*)^m g_q^l}{m!l!} (b^\dagger a)^m (ba^\dagger)^l. \tag{S2.3}$$

We substitute Eq. (S2.3) into Eq. (S2.2), obtaining:

$$\rho_{\text{tot}}^{(f)} = e^{|g_q|^2} \sum_{n,n'=0}^{\infty} \rho_{\text{ph}}(n,n') \sum_{m,l=0}^{\infty} \sum_{j,k=0}^{\infty} \frac{(-g_q^*)^m g_q^l}{m!l!} \frac{(-g_q)^j g_q^{*k}}{j!k!} a^m (a^\dagger)^l |n\rangle \langle n'| a^k (a^\dagger)^j$$
$$\otimes |E_0 + (m-l)\hbar\omega\rangle\langle E_0 + (j-k)\hbar\omega|, \tag{S2.4}$$

where $\rho_{\text{ph}}(n,n') = \langle n|\rho_{\text{ph}}|n'\rangle$ is the element of the photonic density matrix in the number basis. Using

$$\begin{cases} a^m (a^\dagger)^l |n\rangle = \sqrt{\frac{(n+l)!}{n!}} \sqrt{\frac{(n+l)!}{(n+l-m)!}} |n+l-m\rangle \\ \langle n'| a^k (a^\dagger)^j = \sqrt{\frac{(n'+k)!}{n'!}} \sqrt{\frac{(n'+k)!}{(n'+k-j)!}} \langle n'+k-j| \end{cases}, \tag{S2.5}$$

we find the total density matrix in terms of entangled electron–photon states

$$\rho_{\text{tot}}^{(f)} = \sum_{n,n'=0}^{\infty} \rho_{\text{ph}}(n,n') |\Psi_{\text{el-ph}}^n\rangle\langle \Psi_{\text{el-ph}}^{n'}|, \tag{S2.6}$$

where $\left|\Psi_{\text{el-ph}}^{(n)}\right\rangle = \sum_{p=-\infty}^{\infty} C_p^n |E_0 - p\hbar\omega\rangle|n+p\rangle$, and the coefficients are given by:

$$C_p^n = e^{i\varphi_g p} \sum_{l=\max\{0,p\}}^{\infty} e^{\frac{1}{2}|g_q|^2} \frac{(-1)^{l-p}|g_q|^{2l-p}}{(l-p)!l!} \frac{(n+l)!}{\sqrt{n!(n+p)!}}.$$

We simplify further the expression for $C_p^n$ and get:

$$C_p^n = e^{i\varphi_g p} e^{\frac{1}{2}|g_q|^2} \frac{|g_q|^{|p|}}{|p|!} \begin{cases} \sqrt{\frac{(n+p)!}{n!}} {}_1F_1\left(1+n+p, 1+p, -|g_q|^2\right), \ p > 0 \\ (-1)^{|p|} \sqrt{\frac{n!}{(n-|p|)!}} {}_1F_1\left(1+n, 1+|p|, -|g_q|^2\right), \ p < 0 \end{cases}, \quad (S2.7)$$

where ${}_1F_1$ is the hypergeometric function. An equivalent representation of Eq. (S2.7) appears after the Eq. (4) in the main text.

We note that in the limit of weak interactions $|g_q| \ll 1$, the coefficients $C_p^n$ are simplified. Rearranging the terms, we get

$$C_p^n = e^{i\varphi_g p} \sqrt{\frac{(n+p)!}{n^p n!}} \sum_{l=\max\{0,p\}}^{\infty} e^{\frac{1}{2}|g_q|^2} \frac{(-1)^{l-p}|g_q|^{2l-p}\sqrt{n^{2l-p}}}{(l-p)! l!} \frac{(n+l)!}{n^{l-p}(n+p)!}. \quad (S2.8)$$

For small $|g_q| \ll 1$, the dominant contributions come from large $n$. In this limit, we obtain

$$\frac{(n+l)!}{n^{l-p}(n+p)!} \to 1, \ \sqrt{\frac{(n+p)!}{n^p n!}} \to 1, \ e^{\frac{1}{2}|g_q|^2} \to 1,$$

such that

$$C_p^n = e^{i\varphi_g p} \sum_{l=\max\{0,p\}}^{\infty} \frac{(-1)^{l-p}}{(l-p)! l!} \left(|g_q|\sqrt{n}\right)^{2l-p} = e^{i\varphi_g p} J_p(2|g_q|\sqrt{n}). \quad (S2.9)$$

Coherent light: the absence of electron–photon entanglement for strong coherent light

In the case of coherent light, we have the following photon density matrix:

$$\rho_{\text{ph}}(n, n') = e^{-|\alpha|^2} \frac{\alpha^n}{\sqrt{n!}} \frac{\alpha^{*n'}}{\sqrt{n'!}}, \quad (S2.10)$$

Since the density matrix in Eq. (S2.10) is separable in $n$ and $n'$, we can deduce that the total density matrix is a pure state:

$$\rho_{\text{tot}}^{(f)} = |\Psi_\alpha\rangle\langle\Psi_\alpha|, \quad (S2.11)$$

where

$$|\Psi_\alpha\rangle = \sum_{p=-\infty}^{\infty} \left(\sum_{n=0}^{\infty} e^{-\frac{1}{2}|\alpha|^2} \frac{\alpha^n}{\sqrt{n!}} |n+p\rangle\right) C_p^n |E_0 - p\hbar\omega\rangle. \quad (S2.12)$$

Hence, we showed that the interaction with coherent light leads to the completely pure state of the joint density matrix. However, Eq. (S2.12) shows that in the general case, there is entanglement between the electron and photon parts of the wavefunction, i.e., $|\Psi_\alpha\rangle$ cannot be decomposed into a tensor product of the two subsystem states. However, in the limit of small $|g_q|$, we can substitute Eq. (S2.9) and get

$$|\Psi_\alpha\rangle = \sum_{p=-\infty}^{\infty} \left(\sum_{n=0}^{\infty} e^{-\frac{|\alpha|^2}{2}} \frac{\alpha^n}{\sqrt{n!}} |n+p\rangle\right) e^{i\varphi_g p} J_p(2|g_q|\sqrt{n}) |E_0 - p\hbar\omega\rangle. \quad (S2.13)$$

For large $|\alpha|^2 \gg 1$, the distribution is narrowly peaked around $n \approx |\alpha|^2$, which allows us to write to a good approximation

$$|\Psi_\alpha\rangle = \sum_{p=-\infty}^{\infty} e^{i\varphi_g p} J_p(2|g_q\alpha|) |E_0 - p\hbar\omega\rangle \otimes |\alpha\rangle, \quad (S2.14)$$

showing that in the semiclassical limit, the final joint electron–photon state is both pure and separable, i.e., not entangled.

Thermal light: the emergence of electron–photon entanglement

The density matrix of the thermal state has the following diagonal form:

$$\rho_{\text{ph}}(n, n') = \frac{1}{\langle n \rangle + 1} \left( \frac{\langle n \rangle}{\langle n \rangle + 1} \right)^n \delta_{n,n'}. \tag{S2.15}$$

Substituting the thermal density matrix Eq. (S2.15) into Eq. (S2.6), we get:

$$\rho_{\text{tot}} = \sum_{n=0}^{\infty} \frac{1}{\langle n \rangle + 1} \left( \frac{\langle n \rangle}{\langle n \rangle + 1} \right)^n |\Psi_{\text{e-ph}}^{(n)}\rangle\langle\Psi_{\text{e-ph}}^{(n)}|, \tag{S2.16}$$

The state Eq. (S2.16) is not a pure state. Moreover, it cannot be separated into a tensor product of electron and light density matrices, i.e., it could contain electron–photon entanglement.

**S2.3 The electron density matrix following the Q-PINEM interaction**

In this section, we calculate the electron density matrix after a Q-PINEM interaction. The electron density matrix after the interaction equals to:

$$\rho_{\text{el}} = \text{Tr}_{\text{ph}}\left[S\rho_{\text{tot}}^{(i)}S^\dagger\right], \tag{S2.17}$$

where $\text{Tr}_{\text{ph}}$ is the trace with respect to the photonic degrees of freedom. We apply the photon trace-out on the total density matrix Eq. (S2.6-7), yielding the electron density matrix:

$$\rho_{\text{el}}^{(f)} = \sum_{k=-\infty}^{\infty} \sum_{k'=-\infty}^{\infty} \underbrace{\left[\sum_{n=0}^{\infty} \rho_{\text{ph}}(n+k, n+k') D_k^n D_{k'}^{n*}\right]}_{\rho_{\text{el}}^{(f)}(k,k')} |E_0 + k\hbar\omega\rangle\langle E_0 + k'\hbar\omega|, \tag{S2.18}$$

where

$$D_k^n = \frac{e^{\frac{|g_q|^2}{2}}}{\sqrt{n!}} e^{-i\cdot\arg(g_q)k} \sum_{m=\max\{0,k\}}^{\infty} \frac{(n+m)!(-1)^m |g_q|^{2m-k}}{\sqrt{(n+k)!}\, m!(m-k)!} = C_{-k}^{n+k}. \tag{S2.19}$$

Note that in terms of the continuous free-electron energies $E, E'$ measured in the experiment, the final electron density matrix can be expressed as the convolution of $\rho_{\text{el}}^{(f)}(k,k')$ with $\rho_{\text{el}}^{(i)}(E,E')$. The first part, defined in Eq. (S2.18) is a discrete electron energy comb with peaks that depend on the photon density matrix. The second part is the initial electron density matrix $\rho_{\text{el}}^{(i)}(E,E')$, which includes its incoherent broadening and zero-loss peak from the process of electron photoemission at the tip. Together, we can write:

$$\rho_{\text{el}}^{(f)}(E, E') = \rho_{\text{el}}^{(i)}(E, E') * \sum_{p=-\infty}^{\infty} \sum_{q=-\infty}^{\infty} \rho_{\text{el}}^{(f)}(k, k') \delta(E - k\hbar\omega)\delta(E' - k'\hbar\omega). \tag{S2.20}$$

We use Eqs. (S2.19-20) to calculate the electron density matrix for different photonic states, as shown in Fig. S2 for different amplifier parameters (see also Section S3 for further details).

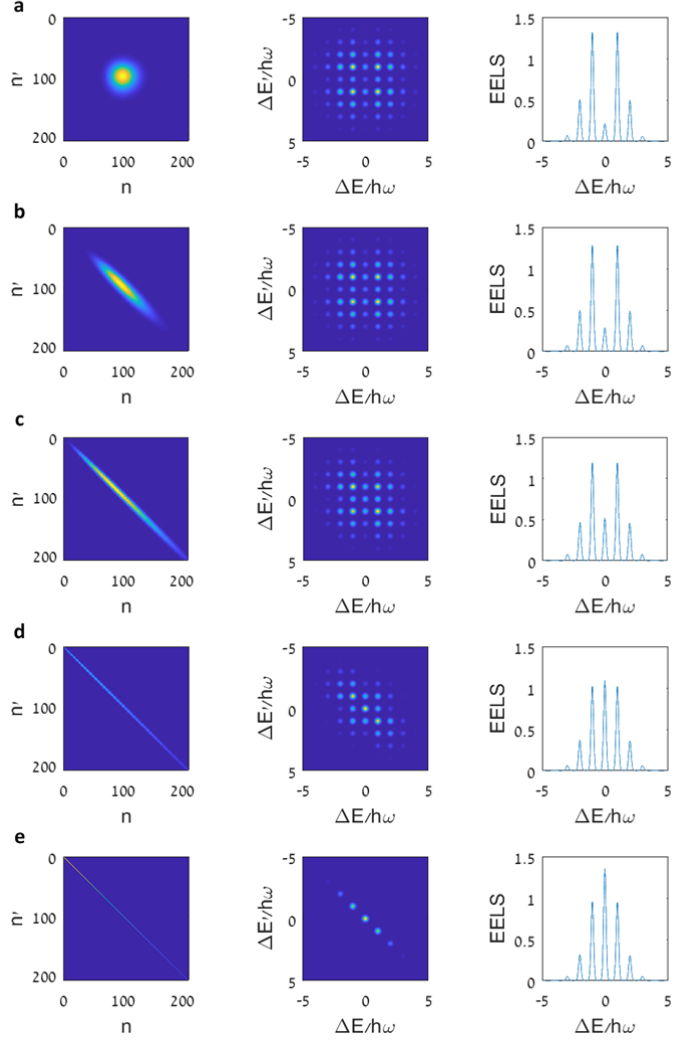

**Figure S2**: **Electron quantum state following a free-electron–light interaction**. Left column: Initial photon density matrix (its absolute values) in the number basis. Middle column: Final electron density matrix (its absolute values) in energy space. Right column: Electron energy spectrum. For all examples, the average number of photons is $\bar{n} = 100$, the quantum coupling constant is $g_q = 0.1$, and the electron initial energy uncertainty is $\Delta E = 0.1\hbar\omega$. **(a)-(e)**: Q-PINEM interaction with an initial photonic state from the amplifier output, ranging between coherent light to thermal light, for a chosen gain of **(a)** $G = 1$, **(b)** $G = 3.37$, **(c)** $G = 11.35$, **(d)** $G = 38.27$ and **(e)** $G = 101$. The photonic quantum state necessary for this calculation is developed in Section S3.5.

### S2.4 Purity of the electron density matrix

This section investigates the purity of the electron state after the interaction to qualitatively estimate the entanglement between the initially pure electron and the light. The purity is defined as follows:

$$\text{purity} = \frac{\text{Tr}[\rho_{\text{el}}^2]}{(\text{Tr}\rho_{\text{el}})^2} = \frac{\sum_{p,q} \rho_{\text{el}}(p,q)\rho_{\text{el}}(q,p)}{1} = \sum_{p,q} |\rho_{\text{el}}(p,q)|^2. \tag{S2.21}$$

Coherent light: purity of the electron density matrix after interaction with strong coherent light

For coherent states, substituting Eq. (S2.10) into Eq. (S2.18), we get the electron density matrix:

$$\rho_{\text{el}}(k,k') = \sum_n \mathcal{D}_k^n \mathcal{D}_{k'}^{n\,*} e^{-|\alpha|^2} \frac{\alpha^{n+k}\alpha^{*n+k'}}{\sqrt{(n+k)!}\sqrt{(n+k')!}} = \sum_n \mathcal{D}_k^n \mathcal{D}_{k'}^{n\,*} \underbrace{e^{-|\alpha|^2}\frac{|\alpha|^{2n}}{n!}}_{p(n)} \frac{n!\,\alpha^k \alpha^{*k'}}{\sqrt{(n+k)!}\sqrt{(n+k')!}}, \quad (S2.22)$$

For strong coherent light $|\alpha|^2 \gg 1$ and $|g_q| \ll 1$, $p(n)$ is narrowly peaked around $n \approx |\alpha|^2$ such that:

$$\rho_{\text{el}}(k,k') \approx \mathcal{D}_k^{|\alpha|^2} \left(\mathcal{D}_{k'}^{|\alpha|^2}\right)^* \frac{\Gamma(|\alpha|^2+1)\alpha^k \alpha^{*k'}}{\sqrt{\Gamma(|\alpha|^2+k+1)\cdot\Gamma(|\alpha|^2+k'+1)}} =$$
$$= \left(\mathcal{D}_k^{|\alpha|^2} \sqrt{\frac{\Gamma(|\alpha|^2+1)}{\Gamma(|\alpha|^2+k+1)}} \alpha^k\right) \cdot \left(\mathcal{D}_k^{|\alpha|^2} \sqrt{\frac{\Gamma(|\alpha|^2+1)}{\Gamma(|\alpha|^2+k'+1)}} \alpha^{k'}\right)^*. \quad (S2.23)$$

Hence, according to Eq. (S2.23), the final electron density matrix is separable:

$$\rho_{\text{el}}(k,k') = \psi_{\text{el}}(k)\psi_{\text{el}}^*(k'), \quad (S2.24)$$

which implies purity of unity because

$$\text{purity} = \sum_{p,q}|\rho_{\text{el}}(p,q)|^2 = \sum_{p,q}\rho_{\text{el}}(p)\rho_{\text{el}}(q) = 1. \quad (S2.25)$$

This is in accordance with Eq. (S2.14), where we saw that the final electron–photon state is pure and separable.

Thermal light: purity of the electron density matrix after interaction

In the case of the interaction with a thermal state, the initial density matrix of the photonic state is given by Eq. (S2.15). Hence, the density matrix of the electron according to Eq. (S2.18) is diagonal:

$$\rho_{\text{el}}(k,k') = \delta_{k,k'} \sum_n |\mathcal{D}_p^n|^2 \frac{1}{\langle n\rangle+1}\left(\frac{\langle n\rangle}{1+\langle n\rangle}\right)^{n+k}, \quad (S2.26)$$

showing that the density matrix is diagonal. According to Eq. (S2.26), the purity equals:

$$\text{purity} = \sum_{p,q}|\rho_{\text{el}}(p,q)|^2 = \sum_k |\rho_{\text{el}}(k,k)|^2 = \sum_k P_k^2, \quad (S2.27)$$

where $P_k$ is the electron energy probability (as given for thermal light in Eq. (S2.32) below). Note that Eq. (S2.27) also holds for free-electron interaction with a Fock state.

In Fig. S3, we plot the purity of the final electron state following an interaction as a function of the interaction strength. In Section S4, the purity of the final electron state is plotted in Fig. S7 as a function of amplifier parameters, showing the transition from coherent to thermal light.

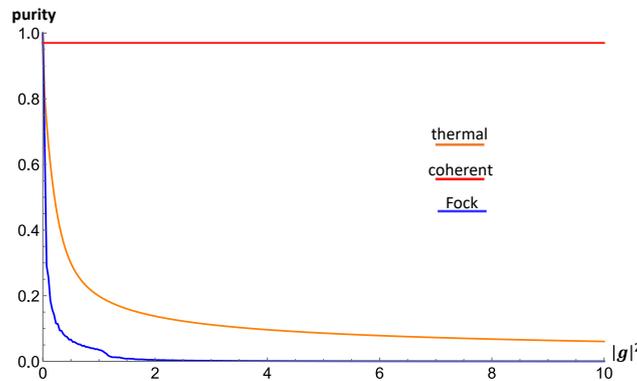

**Figure S3**: Purity of the electron's final state following an interaction with different states of light, as a function of the interaction strength $|g| \equiv |g_q|\sqrt{n}$.

## S2.5 Analysis of the diagonal elements of the total density matrix and quantifying the electron–light entanglement

According to Eq. (S2.4) and according to Ref. S3, the diagonal elements of the total density matrix following the Q-PINEM interaction $P_{nk} \equiv (\langle n| \otimes \langle E_0 + k\hbar\omega|)\rho_{\text{tot}}^{(f)}(|E_0 + k\hbar\omega\rangle \otimes |n\rangle)$ are:

$$P_{nk} = \begin{cases} |g_q|^{2k} e^{|g_q|^2} \frac{(n+k)!}{(k!)^2 n!} \left|{}_1F_1\left(n+k+1, k+1, -|g_q|^2\right)\right|^2 p_{n+k}, & k > 0 \\ |g_q|^{2|k|} e^{|g_q|^2} \frac{n!}{(k!)^2 (n-|k|)!} \left|{}_1F_1\left(n+1, |k|+1, -|g_q|^2\right)\right|^2 p_{n-|k|}, & k < 0 \end{cases}, \quad (S2.28)$$

where $p_n$ is the probability of the light to have $n$ photons (satisfying $p_n = \sum_k P_{nk}$). In the case of large average number of photons $\langle n \rangle \gg 1$ and weak interaction $|g_q| \ll 1$, Eq. (S2.28) can be simplified to:

$$P_{nk} = \left|J_{|k|}(2|g_q|\sqrt{n+k})\right|^2 \cdot p_{n+k}. \quad (S2.29)$$

We use the following formula to quantify the light–electron corrections (plotted in Fig. S4):

$$\text{correlations} \equiv \sum_{n,k} |P_{nk} - p_n P_k|, \quad (S2.30)$$

where $P_k = \sum_n P_{nk}$ is the probability of the electron to absorb $k$ photons. If the joint state has vanishing correlations, it means that the state is separable and does not contain any entanglement between the electron and photons. For correlations > 0, the state could be entangled, and Eq. (S2.30) provides an estimate for how strong the entanglement is. In Fig. S4, we plot the correlations of the final joint state for two cases: interaction with coherent and with thermal light as a function of the interaction strength. As can be seen in the figure, the coherent state interaction creates weak correlations, while for the thermal state interaction, the correlations are strong. This result explains the conclusions in the main text about the role of photon statistics in creating quantum correlations and altering the eventual electron energy spectrum.

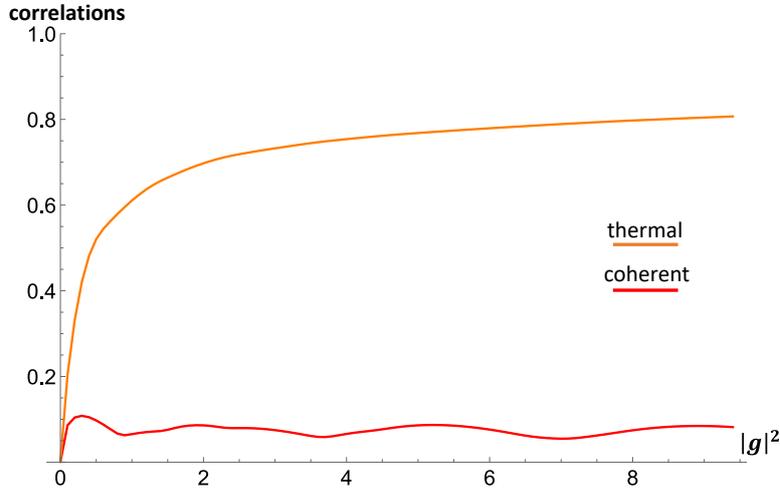

**Figure S4**: **Correlations between the light and the free electron showing their entanglement.** The correlations are plotted for the interaction with coherent and with thermal light. We see that the thermal state creates stronger correlations with the electron than the coherent state. The curves are calculated for $|g_q| = 0.1$, using Eq. (S2.30).

## S2.6 Analysis of the diagonal elements of the electron density matrix and predictions for the measured electron energy spectrum

The diagonal elements of the electron density matrix, according to Eq. (S2.21), equals[S3]:

$$P_k = \begin{cases} \text{Tr}\left[\rho_{\text{ph}} \cdot |g_q|^{2k} e^{|g_q|^2} \frac{\hat{n}!}{(k!)^2(\hat{n}-k)!} \left|{}_1F_1\left(\hat{n}+1, k+1, -|g_q|^2\right)\right|^2\right], & k > 0 \\ \text{Tr}\left[\rho_{\text{ph}} \cdot |g_q|^{2|k|} e^{|g_q|^2} \frac{(\hat{n}+|k|)!}{(|k|!)^2 \hat{n}!} \left|{}_1F_1\left(\hat{n}+|k|+1, |k|+1, -|g_q|^2\right)\right|^2\right], & k < 0 \end{cases}. \quad (S2.31)$$

In the case of $\langle n \rangle \gg 1$ and $|g_q| \ll 1$, we can simplify Eq. (S2.31):

$$P_k = \sum_n |J_k(2|g_q|\sqrt{n+k})|^2 p_{n+k}. \quad (S2.32)$$

The coherent state has the following statistics:

$$p_n = e^{-\langle n \rangle} \frac{\langle n \rangle^n}{n!}, \quad (S2.33)$$

where the average number of photons equals $\langle n \rangle = |\alpha|^2$. Substituting Eq. (S2.33) into Eq. (S2.32), we approximately get for $g_q \ll 1$:

$$P_k \approx |J_{|k|}(2|g|)|^2, \quad (S2.34)$$

where $|g| = |g_q|\sqrt{\langle n \rangle}$.

The thermal statistics has the following form:

$$p_n = \frac{1}{1+\langle n \rangle}\left(\frac{\langle n \rangle}{1+\langle n \rangle}\right)^n. \quad (S2.35)$$

Substituting Eq. (S2.35) into Eq. (S2.32), we get the following electron energy spectrum:

$$P_k = e^{-2|g|^2} I_{|k|}(2|g|^2), \quad (S2.36)$$

where again $|g| = |g_q|\sqrt{\langle n \rangle}$. The electron energy spectrum is calculated according to Eq. (S2.34) and Eq. (S2.36) and plotted in Fig. S5.

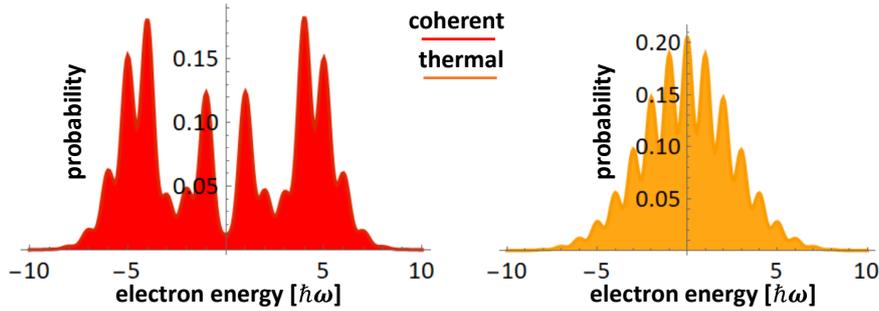

**Figure S5**. **Electron energy spectra for the interactions with coherent and thermal light**, calculated according to Eq. (S2.34) and Eq. (S2.36).

## S2.7 Quantum weak measurement of photonic states using free electrons

In this section, we consider how the electron affects the light during its interaction. We calculate the quantum fidelity, which shows when the electron can be understood as a probe performing quantum weak measurement of the state of light.

We consider the light as our system and the free electron as the measurement device (pointer). The final joint state following the interaction is described by Eqs. (S2.6-7). Our pointer (electron) is measured strongly, i.e., projected on a specific energy value $M = |E'\rangle\langle E'|$, giving the final joint state

$$M\rho_{\text{tot}}^{(f)}M = \rho_{\text{ph}}^{(k)} \otimes |E'\rangle\langle E'|. \quad (S2.37)$$

Therefore, the act of measurement is a detection of a certain $k^{\text{th}}$ electron energy peak. As in the previous sections, we use the index $k = \frac{E'-E}{\hbar\omega}$ to represent the discretized electron energy change, with $k > 0$ corresponding to subtraction of photons from the field. The density matrix of the photonic system $\rho_{\text{ph}}^{(k)}$ following the measurement can be written in a closed-form expression when considering a weak interaction ($|g_q| \ll 1$) using Eq. (S2.9):

$$\rho_{\text{ph}}^{(k)} = \frac{1}{N^{(k)}}\sum_{n=0}^{\infty}\sum_{n'=0}^{\infty}\rho_{\text{ph}}(n+k, n'+k)J_k(2|g_q|\sqrt{n+k})J_k(2|g_q|\sqrt{n'+k})|n\rangle\langle n'|. \quad (S2.38)$$

The normalization factor $N^{(k)}$ is

$$N^{(k)} = \sum_{n=0}^{\infty}\rho_{\text{ph}}(n+k, n+k)J_k^2(2|g_q|\sqrt{n+k}) = \sum_{n=0}^{\infty}p_{n+k}J_k^2(2|g_q|\sqrt{n+k}). \quad (S2.39)$$

To quantify what is considered as a quantum weak measurement, we calculate the fidelity $F_k$ between the state of the light $\rho_{\text{ph}}^{(k)}$ after the measurement and its initial state $\rho_{\text{ph}}$. Weak measurement is defined by fidelity $F_k$ that approaches unity[S4]:

$$F_k = \left[\text{tr}\sqrt{\sqrt{\rho_{\text{ph}}}\rho_{\text{ph}}^{(k)}\sqrt{\rho_{\text{ph}}}}\right]^2 \to 1, \quad (S2.40)$$

In Fig. S6, we plot the fidelities $F_k$ for different quantum optical states (coherent state, thermal state, and Fock state). For a coherent state of many photons with a small $g_q$, the measurement is always weak, whereas for the Fock state, it is always strong, except for the case of post-selection on $k = 0$. For a thermal state, the fidelity is lower than with a coherent state and decays more quickly for larger $|k|$. Interestingly, the fidelity of the measured coherent state is asymmetric around $k = 0$. The reason for this behavior stems from the fundamental difference between photon numbers subtracted and photon number added coherent states[S5], which follow substantially different photon statistics. For photon subtraction ($k > 0$), the state is retained, whereas for photon addition ($k < 0$), the state changes more strongly. Still, both cases undergo a smaller change relative to the interaction with a thermal state. For a thermal state, the fidelity reduces substantially for both photon subtraction and addition in a rather symmetric fashion. Obviously, for the Fock state, adding or subtracting even just one photon completely changes the state, and the fidelity vanishes for all $k \neq 0$.

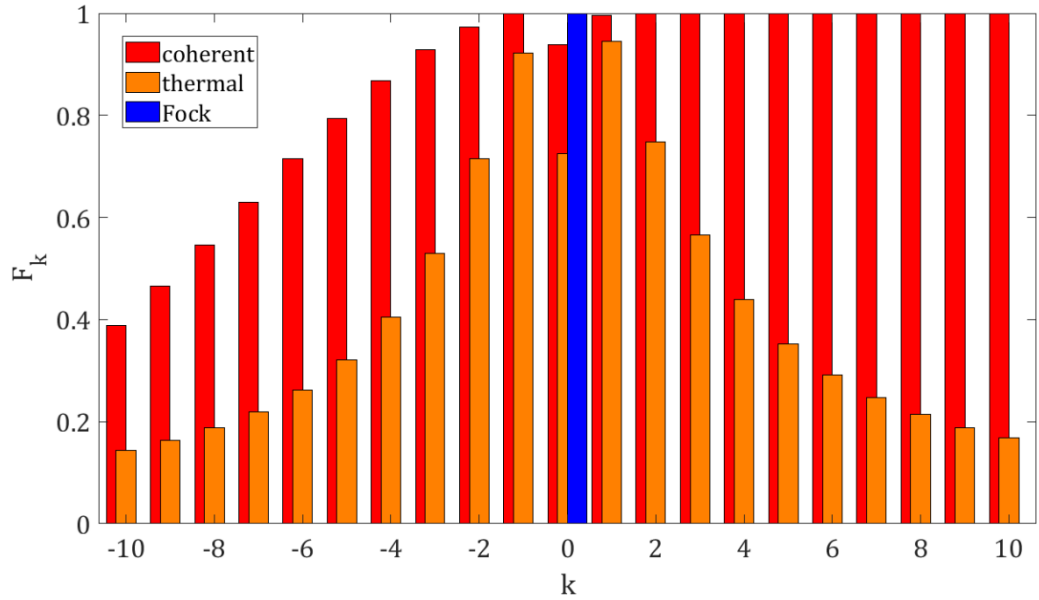

**Figure S6**: **Quantum weak measurement vs. projective measurement. The fidelity quantifies how much the photonic state changes due to the electron interaction.** The curves show the fidelity of the photonic state (the measured system) following the detection of the electron (the measuring pointer) at the $k$-th energy peak, for a coherent state (green), thermal state (red), and a Fock state (blue). All curves are calculated with a mean photon number of $\langle n \rangle = 100$ and $g_q = 0.1$.

# S3. Quantum optical properties of amplifier light: theory and analysis of the amplifier

In this section, we analyze the quantum properties of the light created by an amplifier.

## S3.1. Quantum optical analysis of an amplifier model

To calculate the photon statistics in our experiment, we model our amplifier as a traveling-wave gain medium. This approach is a standard method to analyze a variety of amplifying gain media, such as fiber amplifiers. We denote the input field as $a_{\text{in}}$, the noise field by $b_\mathcal{N}$, and the amplifier gain by $\mathcal{G} > 1$. The output field is then given as[S6]:

$$a_{\text{out}} = \sqrt{\mathcal{G}} a_{\text{in}} + \sqrt{\mathcal{G}-1}\, b_\mathcal{N}^\dagger. \tag{S3.1}$$

This corresponds to an effective two-mode squeezing operation of the form

$$S(r) = \exp[r a_{\text{in}} b_\mathcal{N} - r a_{\text{in}}^\dagger b_\mathcal{N}^\dagger], \tag{S3.2}$$

where the squeeze parameter $r$ satisfies $\sqrt{\mathcal{G}} = \cosh r$. The initial state of the amplifier is modeled as a coherent input seed $|\alpha\rangle$ and a vacuum noise field $|0\rangle$ such that the initial state is $|\Psi_i\rangle = |\alpha, 0\rangle$. Subsequently, the final state satisfies

$$|\Psi_f\rangle = S(r)|\alpha, 0\rangle = S(r) D_{\text{in}}(\alpha)|0,0\rangle, \tag{S3.3}$$

with $D_{\text{in}}(\alpha)$ denoting a displacement operator of the input field mode. We can exchange the order of the displacement and squeeze operator to have[S7]:

$$|\Psi_f\rangle = D_{\text{in}}(\alpha \cosh r) D_\mathcal{N}(-\alpha^* \sinh r) S(r)|0,0\rangle, \tag{S3.4}$$

where now, $D_\mathcal{N}(\alpha)$ is a displacement operator in the noise field.

We need to trace out the noise field if we want to find the state of the output field. Tracing out one mode of the two-mode squeezed coherent state given in Eq. (S3.4) yields a displaced thermal state for the output field, given by the Glauber $P$-function[S8], which is a function of a complex number $\beta$

$$P(\beta) = \frac{1}{\pi(\mathcal{G}-1)} \exp\left[-\frac{(\beta - \sqrt{\mathcal{G}}\alpha)^2}{\mathcal{G}-1}\right]. \tag{S3.5}$$

From the $P$-function we can calculate the probability distribution by integrating over the complex plane $p_n = \int d^2\beta\, P(\beta) e^{-|\beta|^2} |\beta|^{2n}/n!$, yielding

$$p_n = e^{-|\alpha|^2} \frac{1}{\mathcal{G}} \left(1 - \frac{1}{\mathcal{G}}\right)^n L_n\left(-\frac{|\alpha|^2}{\mathcal{G}-1}\right), \tag{S3.6}$$

where $L_n(x)$ is the $n$-th Laguerre polynomial. According to Eq. (S3.6), the photon statistics is described by $\alpha$, the coherent amplitude of the input seed and $\mathcal{G} = \mathcal{G}(|\alpha|^2)$, the gain function of the amplifier, which is defined as the ratio between the output and input power (shown below in Fig. S9d). We use these two parameters, $\alpha$ and $\mathcal{G}$, to characterize the regime of operation of the amplifier. This way, the fitting enables us to extract the amplifier gain curve $\mathcal{G} = \mathcal{G}(|\alpha|^2)$.

For a weak seed $|\alpha|^2 \ll 1$ and $\mathcal{G} \gg 1$, we obtain the expected limit of a thermal state. In the other limit, a coherent-like state with Poissonian statistics is obtained for a saturated amplifier with an amplification $\mathcal{G}$, and an average number of photons of the input seed $\langle n_{\text{in}} \rangle = |\alpha|^2 \gg 1$. In between, we obtain a range of states with super-Poissonian statistics, which span the entire range from thermal to the Poissonian statistics of coherent states. This situation can also be viewed in the following manner: when the amplifier is in the linear regime, the added noise has pure thermal statistics. The noise dominates the output light for a very weak seed. For a larger seed $|\alpha|^2 \gg 1$, the amplifier saturates. The added noise spectral density reduces substantially[S9], and the amplifier noise statistics changes[S10]. However, the signal to noise ratio is very large, so the Poissonian statistics of the amplified signal dominates, and a coherent-like state is obtained, i.e., the statistics is Poissonian even if no always having the same degree of coherence as an ideal coherent state. In between, the added noise has contributions from the modified amplifier noise statistics and from the Poissonian amplified signal so that the resulting noise exhibits a transition from thermal to Poissonian statistics that is determined by $|\alpha|^2$. We now consider the limiting cases of the coherent-like and the thermal states in more detail.

The limit of amplified coherent light:

Let us consider the case $|\alpha|^2 \gg 1$ and saturated $\mathcal{G}$. In this case, Eq. (S3.6) has Poissonian statistics:

$$p_n \approx e^{-\langle n \rangle} \frac{\langle n \rangle^n}{n!}, \qquad (S3.7)$$

where $\langle n \rangle \approx \mathcal{G} \cdot |\alpha|^2$. This equation completely coincides with the statistics of coherent light given by Eq. (S2.33).

The limit of amplified spontaneous emission (ASE), creating thermal light:

Let us consider a completely different case $|\alpha|^2 \ll 1$ and $\mathcal{G} \gg 1$. In this case, the statistics of Eq. (S3.6) can be simplified in the following way:

$$p_n = \frac{1}{\langle n \rangle + 1} \left( \frac{\langle n \rangle}{\langle n \rangle + 1} \right)^n, \qquad (S3.8)$$

where $\langle n \rangle \approx \mathcal{G} - 1$. This equation completely coincides with the statistics of thermal light given by Eq. (S2.35).

The general case:

We now consider the general case between thermal and coherent light and find the average number of photons and the second-order correlation $g^{(2)}(0)$ for the amplifier:

$$\langle n_{\text{out}} \rangle = \mathcal{G} \cdot \langle n_{\text{in}} \rangle + \mathcal{G} - 1, \qquad (S3.9)$$

$$g^{(2)}(0) = 2 - \left( \frac{\mathcal{G} \cdot \langle n_{\text{in}} \rangle}{\mathcal{G} \cdot \langle n_{\text{in}} \rangle + \mathcal{G} - 1} \right)^2, \qquad (S3.10)$$

In the experiment presented in Fig. 3 from the main text, we kept the output power fixed (using an attenuator for the output light), while we increased the input number of photons. In the case of $\mathcal{G} \gg 1$, we can rewrite $g^{(2)}(0)$:

$$g^{(2)}(0) = 2 - \left( \frac{\langle n_{\text{in}} \rangle}{\langle n_{\text{in}} \rangle + 1 - \mathcal{G}^{-1}} \right)^2 \approx 2 - \left( \frac{\langle n_{\text{in}} \rangle}{\langle n_{\text{in}} \rangle + 1} \right)^2. \qquad (S3.11)$$

The $\langle n_{\text{out}} \rangle$ and $g^{(2)}(0)$ as a function of $\langle n_{\text{in}} \rangle$ are shown in Fig. S7 and in Fig. 3 of the main text.

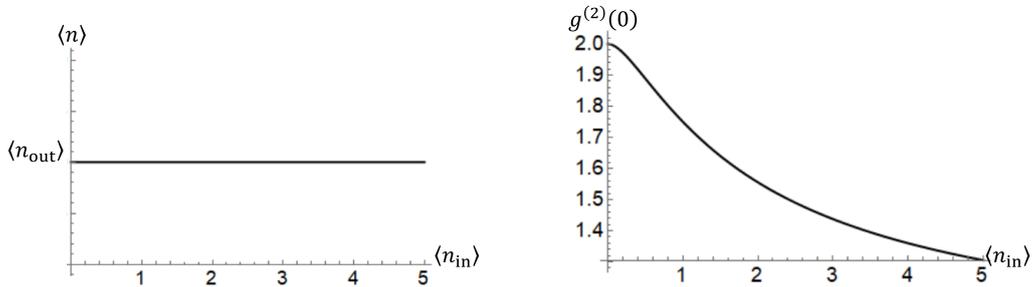

Figure S7: **Transition between coherent and thermal state for the amplifier output** as a function of the input number of photons $\langle n_{in} \rangle$.

We can also calculate the purity of the final electron density matrix, following its Q-PINEM interaction with the amplifier light, as a function of the amplifier gain $\mathcal{G}$. For that, we substitute the general photonic density matrix of the amplifier output (see Section S3.4 below) into Eq. (S2.18). The results of the calculation are depicted in Fig. S8.

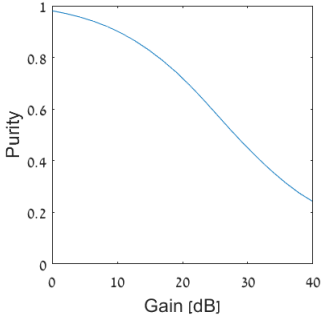

**Figure S8: Purity of the electron quantum state following a Q-PINEM interaction.** The interaction is simulated with an initial photonic state from the amplifier output, ranging from coherent light ($G = 0\ dB$) to thermal light ($G = 40\ dB$). The average number of photons is $\langle n \rangle = 100$, the quantum coupling constant is $g_q = 0.1$, and the initial electron energy uncertainty $\Delta E = 0.1\hbar\omega$.

### S3.2. Analysis of the amplification curve and optical spectra

In addition to the Q-PINEM measurement, we reconstruct the statistics of the amplifier output using the following two optical methods: 1) Using the optical spectra (Fig. S9a); 2) Using the measured amplification curve (Fig. S9d), knowing both the power of the input light and the amplification coefficient $G$. Let us focus on the first method. The typical optical spectra are shown in Fig. S10: it has some level of background noise, amplified noise, and a coherent peak due to the amplified input seed. The amplified noise spectrum has a rectangular shape due to the filter used in the experiment.

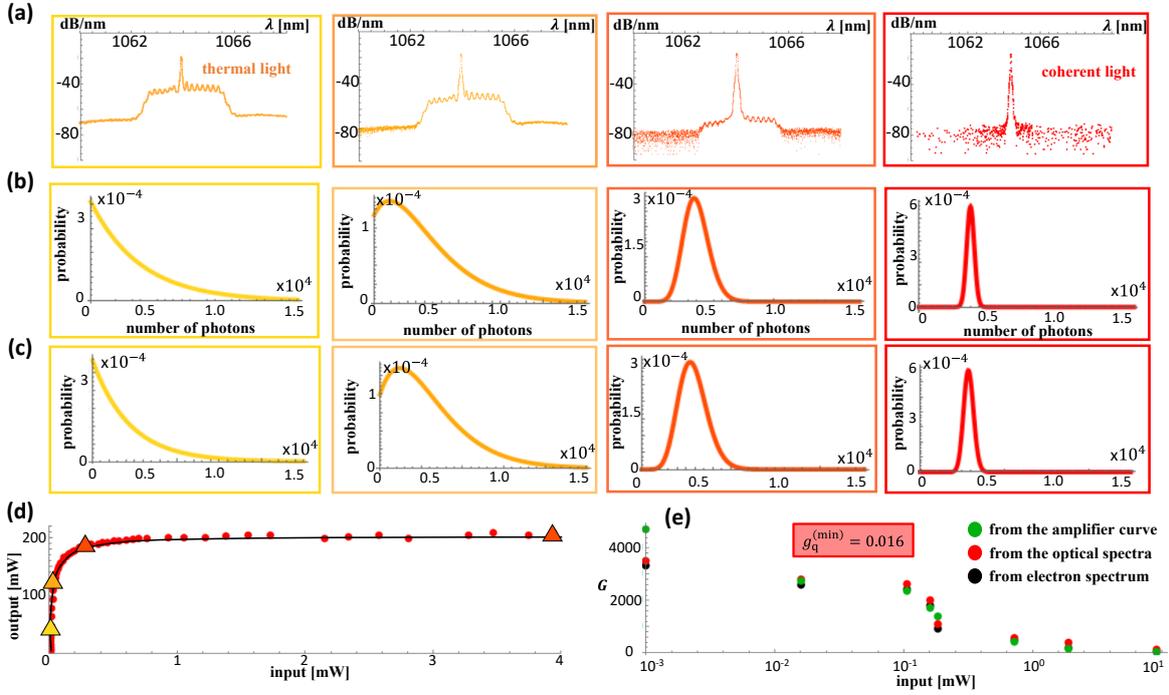

**Fig. S9: Quantum optical properties of the amplifier light.** We compare two methods of analysis of the quantum photon statistics of the amplifier output: **(a)** The optical spectra, from which we extract **(b)** the photon statistics. **(c)** The photon statistics can also be extracted from the measured electron energy spectra following the free-electron–light interaction. We extract the gain G of the amplifier from both experimental measurements and compare with **(d)** a direct measurement of the amplifier gain from input-output measurements. **(e)** The comparison of the three methods shows good agreement and enables us to extract the effective interaction constant $g_q$ up to the coupling efficiency of light into the structure.

To analyze the amplifier statistics from the spectrum, we model the coherent input seed using a Gaussian spectral shape:

$$\frac{d\langle n_{\text{in}}(\lambda)\rangle}{d\lambda} = I_0\, e^{-\frac{(\lambda-\lambda_0)^2}{\sigma^2}}, \tag{S3.12}$$

where $\lambda_0 = 1064$ nm and $\sigma$ are defined by the bandwidth of the laser. The amplifier output signal is the sum of the three components: 1) amplified coherent input Eq. (S3.12); 2) amplified thermal noise, which is created due to the amplified spontaneous emission (ASE); 3) background noise. For $\mathcal{G} \gg 1$, the output can be expressed as

$$\frac{d\langle n_{\text{out}}(\lambda)\rangle}{d\lambda} = \mathcal{G}(\omega)I_0\, e^{-\frac{(\lambda-\lambda_0)^2}{\sigma^2}} + \mathcal{G}(\omega)\cdot \text{noise} + \text{background noise}.$$

According to this model, we can extract $\mathcal{G}(\omega)$ (up to the multiplication constant) from examining the amplification of the noise relative to the unamplified part of the spectrum. Therefore, the value marked as $G$ in Fig. S10 is proportional to the theoretical amplifier gain $\mathcal{G}$; we use it to find the amplification. Furthermore, the height of the Gaussian peak relative to the amplified noise is proportional to the input seed photon number $\langle n_{\text{in}}\rangle$, so we could also use this scaling to find the amplification.

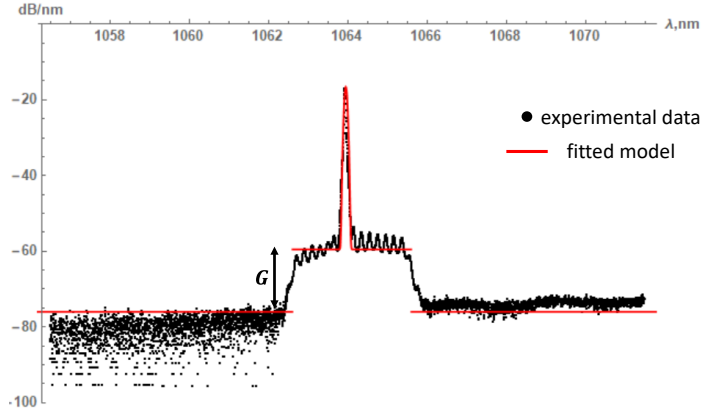

**Figure S10**: **Experimental optical spectrum of the amplifier output and its fit.** The background noise is at -75 dB, the amplified noise is at -60 dB, and the amplified coherent input peak is at -15 dB.

By either analyzing the measured amplification curve (option 1) or the optical spectra (option 2) of the output light, we can estimate the statistical properties of the amplifier output. The results of this analysis are shown in Fig. S9, where green points show the analysis based on the amplifier curve, red points denote the analysis of the optical spectra, and they are both compared with black dots that represent the extracted gain from the electron measurement. The latter is explained in Section S4, detailing the method of analyzing the photon statistics based on the electron energy spectra resulting from the Q-PINEM interaction. The results from this method agree with the photon statistics obtained by the two previous methods (Fig. S9e).

### S3.4. Density matrix description of the amplifier output in the photon number basis

In this section, we derive the photon density matrix of the amplifier output in the photon number basis. According to Eq. (S3.5), the density matrix of the amplified light is:

$$\rho_{\text{ph}} = \int d^2\beta\, \frac{1}{\pi(\mathcal{G}-1)} \exp\left[-\left|\beta-\sqrt{\mathcal{G}}\alpha\right|^2/(\mathcal{G}-1)\right] |\beta\rangle\langle\beta|. \qquad (S3.13)$$

To convert the density matrix to the photon number basis, we need to solve the following integral:

$$\rho_{\text{ph}}(n,n') = \int d^2\beta\, \frac{1}{\pi(\mathcal{G}-1)} \exp\left[-\frac{|\beta-\sqrt{\mathcal{G}}\alpha|^2}{\mathcal{G}-1}\right] e^{-|\beta|^2} \frac{\beta^n}{\sqrt{n!}} \frac{\beta^{*n'}}{\sqrt{n'!}}. \qquad (S3.14)$$

This is a two-dimensional Gaussian integral. The integral Eq. (S3.14) can be evaluated explicitly as:

$$\rho_{\text{ph}}(n,n') = \frac{\alpha^{*n'-n}}{\mathcal{G}} \left(\frac{1}{\sqrt{\mathcal{G}}}\right)^{n'+n} \sum_{m=\max\{0,n-n'\}}^{n} \frac{\sqrt{n!n'!}}{(m+n'-n)!} \frac{(\mathcal{G}-1)^{n-m}}{(n-m)!} \frac{\exp(-|\alpha|^2)|\alpha|^{2m}}{m!}. \qquad (S3.15)$$

Eq. (S3.15) gives the entire photon density matrix of the amplified light. We note that, as expected, the diagonal $\rho_{\text{ph}}(n,n) = p_n$ recovers the statistics of Eq. (S3.6).

# S4. Free-electron interaction with amplified light: Q-PINEM theory and fit to experiments

## S4.1 Applying the Q-PINEM theory for the case of amplifier light

In this section, we revisit derivations from Sections 2 and 3 to explain how our electron–light interaction theory applies to the quantum statistics of photons created from the amplifier theory. Combining these results, we develop the full theory that we later fit to the experimental measurements. According to Eq. (S3.6) and Eq. (S2.32), the electron energy spectra after the interaction with amplified light are described by:

$$P_k = \sum_n |J_{|k|}(2|g_q|\sqrt{n})|^2 e^{-|\alpha|^2} \frac{1}{\mathcal{G}} \left(1 - \frac{1}{\mathcal{G}}\right)^n L_n\left(-\frac{|\alpha|^2}{\mathcal{G}-1}\right), \tag{S4.1}$$

where the spectrum depends on two values: $|\alpha|$ and $\mathcal{G} = \mathcal{G}(|\alpha|^2)$. Eq. (S4.1) generalizes both the interaction with coherent light and with thermal light. In these limiting cases, the electron energy spectra are given by Eq. (S2.34) and Eq. (S2.36), respectively.

To capture the transition between thermal and coherent states, we recall the thermality parameter $r_{\text{th}} \in [0,1)$ that arises from the ratio of the effective number of "thermal photons" to the total number of photons (Section S1.3). Thus, we define $r_{\text{th}}$ as the ratio between the amplifier output without seed to the amplifier output with seed; using Eq. (S3.9), we get:

$$r_{\text{th}} = \frac{\mathcal{G}-1}{\mathcal{G}\cdot|\alpha|^2+\mathcal{G}-1}. \tag{S4.2}$$

We can calculate the electron energy spectrum as a function of $r_{\text{th}}$ using Eq. (S4.1) and Eq. (S4.2) for a fixed amplifier output power, i.e., a fixed $|g| = |g_q|\sqrt{\mathcal{G}\cdot|\alpha|^2 + \mathcal{G} - 1}$. These electron energy spectra are shown in Fig. S1 and are the same as the spectra calculated by the walker theory presented in Section S1, using Eq. (S4.2) and Eqs. (S1.7-8).

## S4.2 Fitting of the experimental data

In this section, we compare the results obtained in the experiment with the theory. We extract the parameters $|\alpha|^2$ and $\mathcal{G}$ by fitting the measured electron energy spectra to Eq. (S4.1). We note that the photon statistics can be more generally extracted from the electron energy spectrum without employing the special case of the amplifier output. This procedure is described in Ref. S3.

Consider the experimental electron energy spectra in Fig. S11a. We want to compare it to the theoretical prediction of Fig. S11b, calculated according to Eq. (S4.1). To do this, we need to consider the fact that the initial electron is not monoenergetic and that there is an interaction between the free electron and the nanostructure, even in the absence of light. This interaction is due to the continuum of modes into which the electron emits spontaneously – which is exactly the effect of electron energy loss spectroscopy (EELS). To take this phenomenon into account empirically, we perform EELS measurements on the nanostructure in the absence of incident light (Fig. S11c). We then use the measured EELS (Fig. S11c) to convolve the results of our single-mode Q-PINEM theory.

Altogether, to compare the theory Fig. S11b with the experimental Fig. S11a, we convolve the theoretical peaks with the experimentally obtained zero-loss peak (Fig. S11c), obtaining the final theoretical spectrum, as in Fig. S11d. To fit the model to the experimental data, we use two fitting parameters from the theory: $a = \eta|g_q|^2\mathcal{G}$ and $b = \eta|g_q|^2\mathcal{G}\langle n\rangle$, where $\eta$ is the coupling efficiency of the light with the silicon-photonic nanostructure, i.e., the ratio between the number of photons that are coupled to the nanostructure and the number of photons emitted from the amplifier. In this manner, we fit the theory and experimental results in all the figures of the main text and extract the corresponding fit parameter values of $\eta|g_q|^2\mathcal{G}$ and $\eta|g_q|^2\mathcal{G}\langle n\rangle$ for the different experiments. Thus, for example, we can reconstruct the photon statistics in Fig. 3 from the main text. Moreover, we can calculate the strength of the quantum coupling constant $\sqrt{\eta}|g_q| \approx 0.016$, extracting it from the fit parameter $a = \eta|g_q|^2\mathcal{G}$, by using the gain $\mathcal{G}$ obtained from the amplification curve described in Section S3.

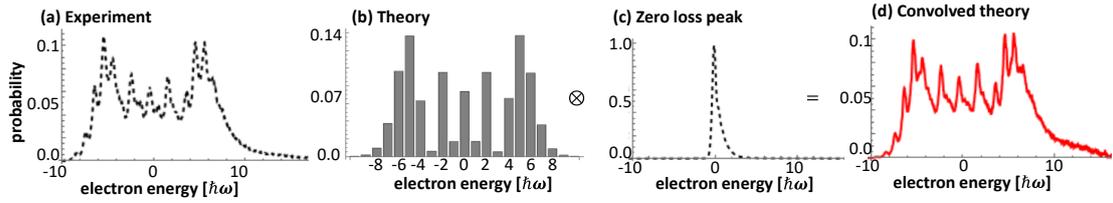

**Figure S11: Fitting theory with experiment. (a)** Experimental electron energy spectrum. **(b)** The electron–light interaction theory (single-mode) according to Eq. (S4.1). **(c)** The experimental energy loss spectrum without an external light illuminaiton. **(d)** Convolution of (b) and (c) provides a very good match with the experimental data.

Fig. S12 presents the comparison between the experimental and the fitted theoretical spectra for different strengths of the electric field, shown for the limits of coherent and thermal light.

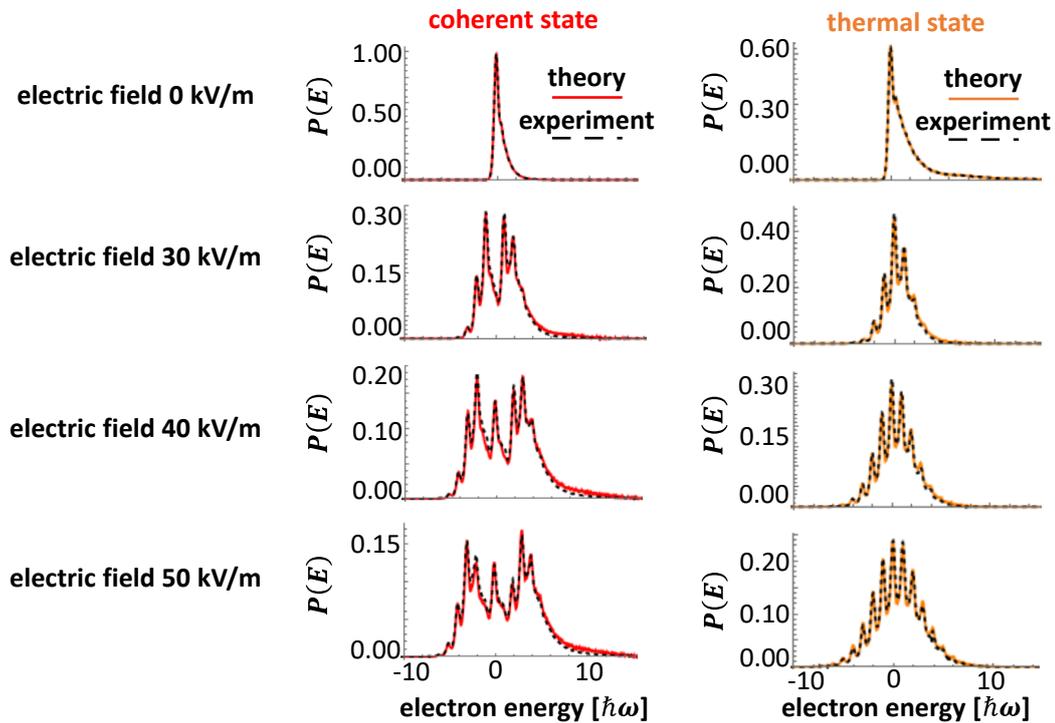

**Figure S12: Electron energy spectra for interactions with coherent and thermal light for selected electric field magnitudes.** Colored curves show the theory, while the dashed black curves show the experimental measurements. The top panels show the electron energy spectrum without any external light illumination, i.e., the electron energy loss during its motion through the structure. The energy loss is due to spontaneous emission into various excitations as in conventional electron microscopy (combination of phonons, plasmons, Cherenkov radiation, and Smith-Purcell radiation). We use the resulting energy loss spectrum as the basic energy peak to be convolved with the discrete prediction of the Q-PINEM theory (or equivalently, the walker theory). Note that the thermal light achieves the same interaction strength as coherent light for the same illumination power (same field amplitude) because it maintains phase-matching in the same manner as coherent light. The differences in the electron energy features are due to the thermal light becoming entangled with the electron in the energy domain.

## S4.3 The asymmetry in the measured electron energy spectra

Figs. 1c,d and Figs. 2a,b,d from the main text show an intrinsic asymmetry in the electron energy spectra. Such asymmetry arises from inelastic scattering of the electron by the structure, including spontaneous emission of a range of optical excitations (usually without phase-matching)[S11,S12]. This behavior contrasts with the symmetry of the rest of the features in the electron energy spectra, which all arise from stimulated emission and absorption. In principle, both spontaneous and stimulated effects can be fully captured by the QPINEM theory when applied to a continuum of optical modes[S11,S13]. However, the single-mode QPINEM theory that we employ (Eqs. (2-4) from the main text and Section S2) does not capture the asymmetry. Thus, we apply an empirical approach to capture the asymmetry of inelastic scattering: we measure the electron energy loss spectrum (EELS)[S11] in the interaction with no external light illumination (Fig. S12), and then we convolve it with the result of the QPINEM theory, showing good agreement with the experimental data.

Despite the free-electron interaction involving a continuum of optical modes, as seen by the asymmetric EELS (Fig. S12), the light excitation only couples to a single mode (the light bandwidth is narrow enough for the single-mode approximation being accurate). Consequently, the free-electron–light interaction acts as a probe of that single mode rather than of the continuum. Importantly, even when the light excitation is broadband and couples to a wider continuum of modes, quasi-phase-matching could enable selective probing of a single mode by tuning the electron velocity.

## S4.4 Mechanisms other than photon statistics that alter the electron energy spectrum

We verify that our observed quantum-to-classical transition does not arise from the lack of spatial/temporal optical coherence. The spectral bandwidth of thermal light in our experiment (4 nm) depicts a longer temporal coherence (900 fs) than the electron–light interaction duration (410 fs). The transverse coherence of light is similarly longer than the electron–light interaction length (84 μm), as it is emitted from a single-mode fiber (6 μm core). Thus, optical coherence is fully maintained even for thermal light. To further strengthen this conclusion, we developed the theory for free-electron interaction with partially coherent light (Section S6). We find that partial optical coherence reduces the efficiency of quasi-phase-matching in a manner that no longer matches with our measurement due to the diminished interaction strength (Fig. S15 below in Section S6). In contrast, we find equal interaction strengths (variance in energy) for both coherent states and thermal states of the same intensity (e.g., white circles in Fig. 1c,d from the main text), implying the same level of optical coherence in both cases. We conclude that the differences in electron energy spectra for different states of light arise *entirely* from differences in photon statistics, leading to quantum decoherence in the energy domain (which can be equally understood in the language of collapse or entanglement).

It is also valuable to compare our measurements of free-electron–quantum-light interactions to previous work in PINEM. Fig. 1c from the main text exactly corresponds to the PINEM theory[S14,S15], showing quantum walk (also called Rabi oscillations in some works) first observed in Ref. S16. The difference is that our experiment observes this effect using CW light rather than a pulsed laser, which is orders of magnitude more intense. Fig. 1d from the main text exactly corresponds to the Q-PINEM theory with thermal light statistics, showing classical random walk, which has a Gaussian distribution. Certain PINEM experiments with coherent laser pulses (e.g., Ref. S17) showed Gaussian-like distributions and may be mistakenly perceived as similar: Importantly, in these experiments, the Gaussian-like electron energy distributions arose from the inhomogeneous temporal field profile along the interaction with the electron (the electron pulse duration was longer than the laser pulse duration), and from the inhomogeneous spatial profile inside the electron beam diameter. In contrast, CW light maintains an average intensity $\langle I(t) \rangle = \langle I \rangle$ that does not depend on time and thus remains homogeneous along the interaction with the electron. Moreover, our electron beam diameter is small enough (Section S8.1) to eliminate spatial inhomogeneities. Hence, the effects we observe here are purely due to photon statistics.

## S5. The phase-matching condition and its analysis

The requirement for a strong interaction between free electrons and extended light fields is matching the electron velocity with the light's phase velocity. This requirement is called the phase-matching condition in electron–light interactions. The phase-matching condition is well-known in classical electrodynamics and forms the basis of efficient dielectric laser accelerators (DLAs)[S18,S19]. This phase-matching condition was recently shown in a PINEM interaction, where the electron is a quantum wavefunction, and the light is a coherent state[S20]. In this section, we describe how the phase-matching condition arises in the PINEM theory and why it is directly applicable in Q-PINEM[S11, S12, S20] for any form of quantum light.

The subsections below present how the phase-matching condition arises straightforwardly from the PINEM theory and then present its generalization to a quasi-phase-matching condition, which appears in our experiment. We use rigorous numerical simulations of the field inside the nanostructure to fully describe the interaction and compare it with the model arising from PINEM theory.

### S5.1. The emergence of phase-matching from the PINEM theory

To find the phase-matching condition in the PINEM theory, let us calculate the strength of the PINEM interaction (see Ref. S20):

$$|g| = \frac{e}{\hbar\omega} \left| \int E_z(z) e^{-i\frac{\omega}{v}z} dz \right|. \tag{S5.1}$$

The field $E_z(z)$ in the structure is given by:

$$E_z(z) = E_0 e^{ik_z z}, \tag{S5.2}$$

where $k_z = k_0 n \cos\theta$ is the projection of the wavevector on the z axis, $k_0 = 2\pi/\lambda_0$ is the wavenumber of the mode, and $\lambda_0$ is the wavelength. Substituting Eq. (S5.2) into Eq. (S5.1), we get:

$$|g| = \frac{e|E_0|}{\hbar\omega} \left| \int_{-L/2}^{+L/2} \exp(ik_0(n\cos\theta - \beta^{-1})z)\, dz \right|, \tag{S5.3}$$

where $n$ is an effective refractive index, $\theta$ is the angle between the electron velocity (i.e., the z-axis) and the wavevector of the light, $\beta = v/c$ is the normalized velocity of the free electron, and $L$ is an effective interaction length.

The integral in Eq. (S5.3) yields:

$$|g| = \frac{e|E_0|}{\hbar\omega} L \left| \mathrm{sinc}\left( \frac{k_0 L}{2}(n\cos\theta - \beta^{-1}) \right) \right|, \tag{S.5.4}$$

The phase-matching condition then reads:

$$n\cos\theta - \beta^{-1} = 0, \tag{S5.5}$$

and the phase-matched value of $|g|$ equals to:

$$|g|^{(\mathrm{max})} = \frac{e|E_0|}{\hbar\omega} L. \tag{S5.6}$$

### S5.2. Quasi-phase-matching in periodic structures: the inverse Smith–Purcell effect

The previous subsection can be understood as an implementation of the inverse Cherenkov effect, as shown in Ref. S20. Our experiment can be explained in an analogous manner as an implementation of the inverse Smith–Purcell effect.

In a periodic structure, the condition for a strong interaction of a free electron with light is called the quasi-phase-matching condition (in analogy to periodic-poled crystals in nonlinear optics). As with other forms of free-electron–light phase-matching, this condition guarantees that the electron velocity is equal to the phase velocity of the optical mode of the light in the structure. Hence, the interaction strength is sensitive to the electron velocity and allows us to selectively couple the electron to

(approximately) a *single mode of light*. Moreover, the quasi-phase-matching condition not only enhances the interaction with a chosen single mode but also suppresses the interaction with the other light modes. This way, we could choose a specific light wavelength by tuning the velocity of the electron (Section S5.4).

In the presence of a periodic structure with periodicity $\Lambda$, the field in Eq. (S5.1) is:

$$E_z(z) = \sum_m E_m e^{i\left(k_0 \cos\theta + \frac{2\pi}{\Lambda}m\right)z}, \tag{S5.7}$$

where $m$ is the Fourier (diffraction) order of the mode with amplitude $E_m$. Then, the interaction strength becomes

$$|g| = \frac{e}{\hbar\omega}\left|\sum_m E_m \int_{-\frac{L}{2}}^{+\frac{L}{2}} \exp\left(i\left(k_0 \cos\theta + \frac{2\pi}{\Lambda}m - \frac{\omega}{v}\right)z\right)dz\right|. \tag{S5.8}$$

Usually, only one diffraction order $m$ is phase matched, hence the coupling is simplified to

$$|g| = \frac{e|E_m|L}{\hbar\omega}\left|\text{sinc}\,\pi L\left(\frac{1}{\lambda_0}\cos\theta + \frac{1}{\Lambda}m - \frac{1}{\lambda_0\beta}\right)\right|, \tag{S5.9}$$

where we substituted $\lambda_0 = 2\pi/k_0$. Then, the phase-matching condition becomes

$$\lambda_0 = \frac{\Lambda}{m}(\beta^{-1} - \cos\theta), \tag{S5.10}$$

which is exactly the Smith-Purcell dispersion relation[S21], describing the spontaneous emission of photons by electrons near a periodic structure. Therefore, our experiment implements the inverse Smith–Purcell effect, which was described theoretically with electron wavefunctions[S22,S23], but so far never showed with CW light nor with non-Poissonian driving light.

### S5.3. The analysis of experimental data

In this part of the section, we discuss the experimental verification of the phase-matching condition. Using the method suggested in Section S4 for each spectrum, we extract the coupling strength $|g|$ as a function of electron velocity. To analyze the phase-matching, we compared these experimental measurements with Eq. (S5.9). We substitute $\lambda_0 = 1064$ nm, $\Lambda = 733$ nm, $m = 1$, and $\theta = 90°$. Using the measured electron energy spectra as a function of electron velocity for interactions with coherent light (Fig. 5a from the main text, also Fig. S13 below), we fit the theory to the experimental results and find the effective interaction length with the structure: $L \approx 56$ μm.

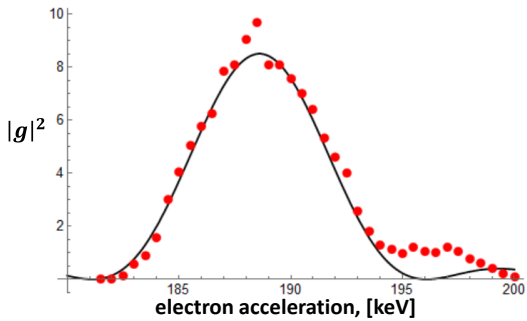

**Figure S13: Comparison between the experiment and theoretical prediction of the phase-matching condition.** The figure shows the coupling strength $|g|^2$ as a function of electron acceleration, calculated according to Eq. (S5.9) (black curve) and extracted from the experiment (red dots).

The rigorous numerical simulation of the electromagnetic fields inside the structure gives a good agreement with the simple model described by Eq. (S5.9). Although the experimental results show an asymmetric behavior that partially deviates from the theory for high electron acceleration, they generally follow the same trend and have the same quasi-phase-matching condition. A more advanced theory could consider the transverse Gaussian profile and defects of the structure. Such theory can potentially provide a more precise fit capturing the asymmetric behavior of the data and the deviation from the theory for high electron velocities.

To summarize this section, we note that the idea of phase-matching between the electron wavefunction and the light wave can lead to strong coupling between free electrons and light. The phase-matching is sensitive to the phase velocity of the mode of light and to the electron velocity, and thus, it allows us to couple exactly to a *single mode of light*. The phase-matching condition not only enhances the interaction with a chosen single mode but also diminishes the interaction with all the other light modes. In this manner, we can selectively couple to a selected light mode by tuning the velocity of the electron. The practical resolution by which the phase-matching condition enables to isolate a single mode is determined by the extent and duration of the interaction (in our case, it is 84 μm, which limits our wavelength resolution to about 8 nm).

**S5.4. Detection of quantum light by using free electrons: prospects for extremely high resolution in both time and frequency**

The quasi-phase-matching shown in Fig. S13 and also in Fig. 5 from the main text could provide a useful capability for quantum-optical detection: resolving the quantum photon statistics of an individual mode (single frequency) out of a multi-mode (or a spectrally broad) pulse of quantum light. By scanning over the electron velocity, as in Fig. 5 from the main text, we can select a particular mode with which the electron satisfies quasi-phase-matching. Then the interaction could potentially isolate the photon statistics of that mode. The spectral resolution improves for quasi-phase-matching of longer extent and duration (see Fig. S14). The maximum wavelength range that can be probed depends on the bandwidth of the nanostructure's spectral response, which is 8 nm in our case (Fig. 5d inset from the main text) and can be significantly wider (as already shown in previous work, e.g., Ref. S24).

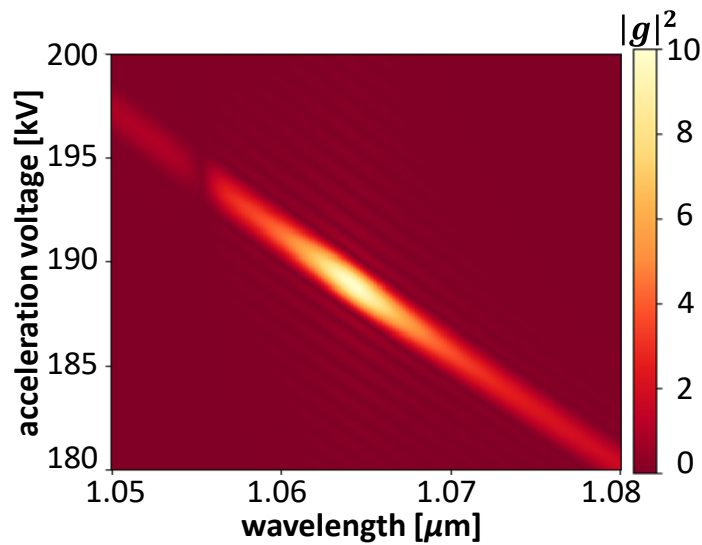

**Figure S14: Phase-matching with a longer structure.** Simulation of 5-times-longer structure, 420 μm, demonstrating the concept of wavelength-selective interaction: controlling the acceleration voltage to enable probing the quantum statistics of photons at the desired frequency. This capability is especially attractive for quantum light of wide bandwidth or ultrafast pulses of quantum light.

Another advantage of free-electron-based quantum-optical detection is the prospects of reaching time resolutions only limited by the electron pulse duration[23]; currently, tens-hundreds of fs. These time resolutions are already better than all existing mechanisms of quantum light detection, which are limited by electronic timescales[S25]. Future work can improve the time resolution to attosecond timescales using attosecond electron combs[S26-S29]. Our free-electron-based approach can also be attractive for wide-bandwidth quantum-optical detection[S30,S31], and potentially lift the bandwidth limitation from applications such as continuous-variable quantum information processing[S30]. As with all other components and detectors in quantum optics, the future success of our detection technique depends on high efficiency (low loss) manipulation of light; specifically, efficient coupling into the nanostructure that performs the electron–light interaction. High-efficiency coupling of light into silicon-photonic nanostructures is a topic of intense investigation and is already available as on-chip technology[S32].

# S6. The role of optical coherence in Q-PINEM interactions

In this section, we consider the model of interaction of free electrons with partially coherent light and show the effect of the partial optical coherence on the electron energy spectra. We emphasize the distinct differences between effects of partial coherence and the effects observed in our experiments due to super-Poissonian and thermal light statistics.

We consider classical coherent light. According to Eq. (S5.1), the coupling between light and free electrons equals to:

$$g = |g|e^{i\phi} = \frac{e}{\hbar\omega}\int E_z(z)e^{-i\frac{\omega}{v}z}dz, \quad (S6.1)$$

where $E_z(z)$ is the classical electric field along the $z$ axis. To model optically incoherent light, we consider consecutive interactions with $N$ coherent fields, each with coherence length $l_c = L/N$ ($L$ is the total interaction length) and a constant phase $\phi_i$ over this interval, where each phase is uniformly distributed in $[0,2\pi]$. This simple model gives the following coupling strength:

$$g = \frac{|g|}{N}\sum_{i=1}^{N} e^{i\phi_i}, \quad (S6.2)$$

The results of the numerical calculations of the electron energy spectra for different $N$ values (i.e., different optical coherence lengths $l_c$, decreasing for larger $N$) are displayed in Fig. S15.

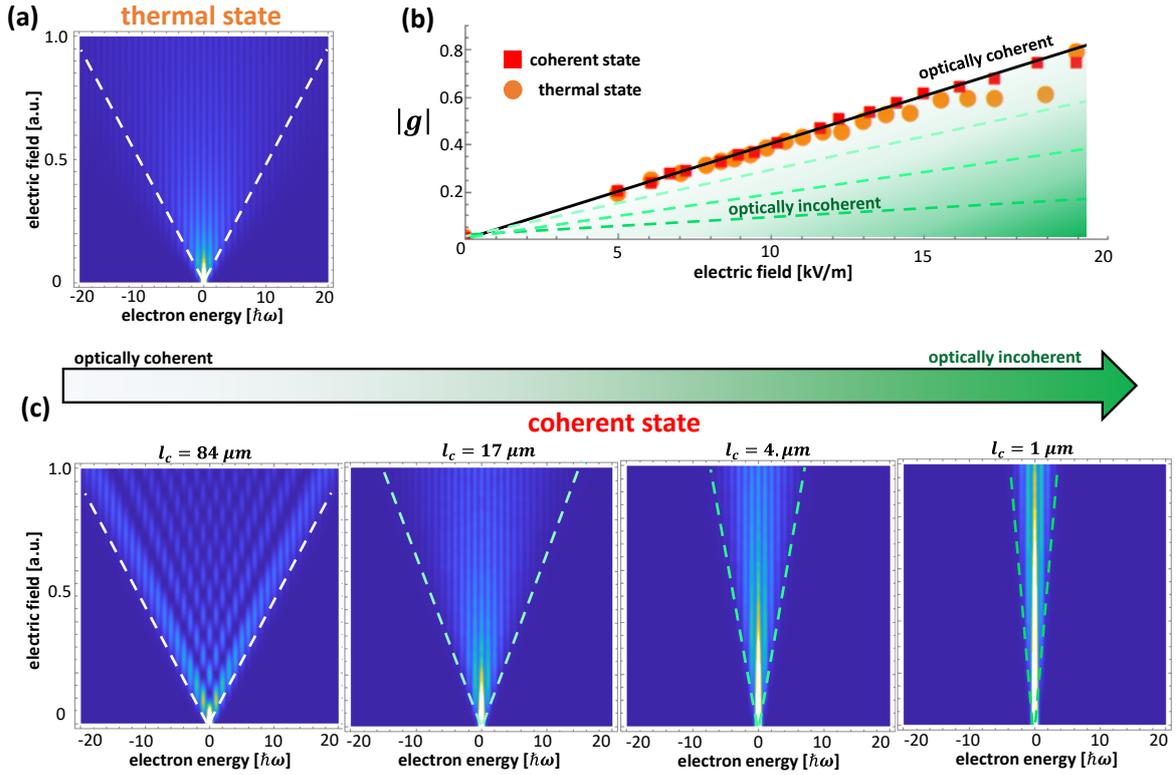

**Figure. S15: The interaction of an electron with optically incoherent light**. (a) Electron energy spectrum as a function of electric field for thermal light. (b) Experimental values of $|g|$ (which is proportional to the electron energy spread after the interaction) as a function of the electric field. It shows that the experimental result obtained for the thermal and coherent light states have the same slope. (c) Electron energy spectra as a function of electric field magnitude for coherent-state light and for different degrees of optical coherence (i.e., optical coherence lengths $l_c$). The energy spread decreases for shorter optical coherence lengths. The experimental results in (b) demonstrate that we have optically coherent light in both thermal and coherent interactions since the thermal and coherent light states have the same energy spread. In the simulations of this figure, the effective interaction length $L$ is assumed to be 84 µm long.

To conclude, we see that for smaller optical coherence of light (i.e., smaller coherence length $l_c$) the energy spread of the electron after the interaction also becomes smaller. This is equivalent, in the quantum walk picture, to a quantum walk in a temporally disordered medium, resulting in diffusion scaling weaker with $N$, until eventual localization emerges (no energy spread). We can use the slope of the energy spread as the function of $|g|$ to identify the degree of optical decoherence. When applying this analysis to our experiment, we find that our data can be explained by fully coherent light, with no need for corrections due to partially incoherent light. Importantly, this section shows that one cannot explain free-electron interactions with thermal light using optical decoherence – these are two completely separated effects (one associated with phase fluctuations in the light, and the other with photon statistics – or intensity fluctuations). In our experiment, the thermal state had full optical coherence in both time (900 fs) and space (assumed perfect since it is emitted from a single mode fiber). These values are larger than the corresponding interaction duration (410 fs) and structure length (84 μm).

## S7. The interaction of classical point electrons with quantum light

In this section, we derive the interaction between a point free electron and quantum light for the case of coherent and thermal states of light. Here we should note that we use a rather exotic assumption: while we treat an electron as a point particle (completely ignoring its wave nature), we consider light as a quantum object. With this purpose, we calculate the electron energy spectrum for interaction with a classical electromagnetic field (i.e., for coherent states of light) and for the interaction with thermal light using the Glauber $P$-function. We use this derivation to compare with our fully quantum analysis (main text Fig. 3), showing that the quantum theory is necessary in all cases.

Coherent light can be described by a classical electromagnetic field $E_z(t)$. Thus, the interaction of the coherent light with an electron can be described by the following equation of motion:

$$m_e \frac{dv}{dt} = -e \cdot E_z(t), \quad (S7.1)$$

where $m_e$ is the mass of the electron, $e$ is the elementary charge, and $v$ is the electron velocity. The magnetic term in Eq. (S7.1) is negligible in our conditions. We assume that the initial electron position is uniformly distributed over a single optical cycle of the electromagnetic field. Within these assumptions, it can be shown that the probability distribution over the energy of the electron after the interaction is (detailed derivation can be found in Ref. S20):

$$P_{\text{coherent}}(\Delta E, \alpha) = \frac{1}{\pi\sqrt{4|g|^2(\hbar\omega)^2 - (\Delta E)^2}}, \quad (S7.2)$$

where $\Delta E$ is the electron energy shift, $|g|^2 = |g_q|^2 \cdot \langle n \rangle = |g_q|^2 |\alpha|^2$ is the number of photons in the coherent light, and $\omega$ is the frequency of the light. The formula in Eq. (S7.2) exactly corresponds to the acceleration in the classical theory of DLA. Fig. S16 (left) compares the classical theory with the quantum (PINEM) theory, showing that they do not match and that only the quantum theory matches with the experimental data.

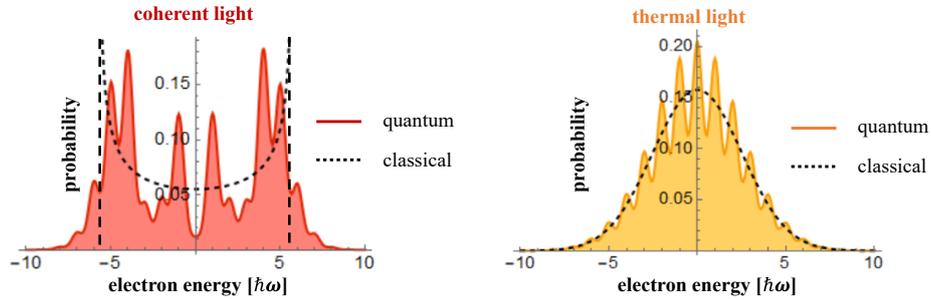

**Figure S16: Comparison between the quantum and classical theory of the PINEM and Q-PINEM interaction.** The figure shows the quantum and classical descriptions of the interaction with coherent (left) light and thermal (right) light, respectively.

We now calculate the interaction of a classical point electron with thermal light. The density matrix of thermal light according to Ref. S33 can be described by the Glauber function:

$$\rho_{\text{thermal}} = \int P(\alpha)|\alpha\rangle\langle\alpha|d^2\alpha, \quad (S7.3)$$

where $|\alpha\rangle$ is the coherent state of light, $P(\alpha) = \frac{1}{\pi\langle n\rangle}e^{-|\alpha|^2/\langle n\rangle}$ is the Glauber function of the thermal light, and $\langle n \rangle$ is the average number of photons. The resulting probability distribution of the electron energy shift equals to:

$$P_{\text{thermal}}(\Delta E) = \int P_{\text{coherent}}(\Delta E, \alpha) \cdot P(\alpha) \, d^2\alpha. \quad (S7.4)$$

Substituting Eq. (S7.3) into Eq. (S7.4), we get:

$$P_{\text{thermal}}(\Delta E) = \frac{1}{2\sqrt{\pi}|g|\hbar\omega} e^{-\left(\frac{\Delta E}{2|g|\hbar\omega}\right)^2}, \tag{S7.5}$$

where $|g|^2 = |g_q|^2 \cdot \langle n \rangle$. Fig. S16 (right) compares the classical theory and the quantum (Q-PINEM) theory, showing that they do not match in the thermal case either.

To summarize, this section presents an attempt at an alternative theory to explain the measurements. This alternative theory does not match with our data, which helps eliminate an explanation of our experiment as a "collapse" or "measurement" of the electron in space. If the thermal light could "measure" ("collapse") the position of the electron, i.e., localize the electron in the space, then such a situation would correspond to the classical theory that we consider in this section. Instead, we find that the electron "collapse" only occurs in the energy domain and that such a quantum theory is the one that correctly matches the measured data. Therefore, our experiment demonstrates that we work in a quantum regime in which the electron wavefunction has a long coherence length (much larger than the wavelength of the light) and a long coherence duration (much longer than the optical cycle).

## S8. The experimental setup

The experimental apparatus is described in the subsections below: the electron microscope, the light system (laser and amplifier), the fabrication, and inverse design of our silicon-photonic nanostructure.

### S8.1. The electron microscope and light source

The experiments were performed using a system based on a JEOL JEM-2100 Plus TEM equipped with a Gatan GIF system and operating using a $LaB_6$ electron filament in thermal emission mode. The TEM was modified to enable coupling light into the sample (described schematically in Fig. 3 from the main text), used previously for ultrafast TEM experiments[S20,S34].

Data acquisition is done in the converged beam electron diffraction (CBED) mode. The electron beam diameter in the focal plane was 30 nm with a convergence angle of 0.3 mrad and a 0.6 eV FWHM zero-loss energy width. We aligned the nanostructure channel to the electron trajectory using a double-tilt holder (Mel-Build Hata Holder) with a custom cartridge to avoid shadowing the optical beam.

A CW-driven distributed feedback (DFB) laser (QLD106p-64D0) emitting at 1060 nm served as a seed for a two-stage Yb fiber amplifier with a 4 nm filter between the two stages. The laser spot size was focused with a cylindrical lens to 51 μm FWHM along the electron trajectory and 8.5 μm FWHM in the direction perpendicular to the electron and to the nanostructure channel.

### S8.2. Fabrication of the silicon-photonic nanostructure

The silicon-photonic nanostructure was fabricated by electron beam lithography (100 keV) and cryogenic reactive-ion etching of 1-5 Ωcm phosphorus-doped silicon to a depth of 2.8±0.1 μm[S35]. The surrounding substrate was etched away to form a 30-μm-high mesa, providing clearance for both electron beam and light beam.

### S8.3. Photonic inverse design of the silicon-photonic nanostructure

We used an open-source Python package based on a 2D-FDFD simulation[S36]. The structure was optimized over a 5-μm-wide design region in the $xz$-plane, containing a 250-nm-wide vacuum channel in the center for the electrons to propagate through. Periodic boundaries were applied along the $z$-direction, enforcing the periodicity of 733 nm, and perfectly matched layers were defined along $x$. A transverse-magnetic plane wave was excited from one side, and the resulting electric field $E_z$ was computed in the center of the channel to find the acceleration gradient, which served as the objective function of the optimization[S37,S38]. The silicon-photonic nanostructure is shown in Fig. S17.

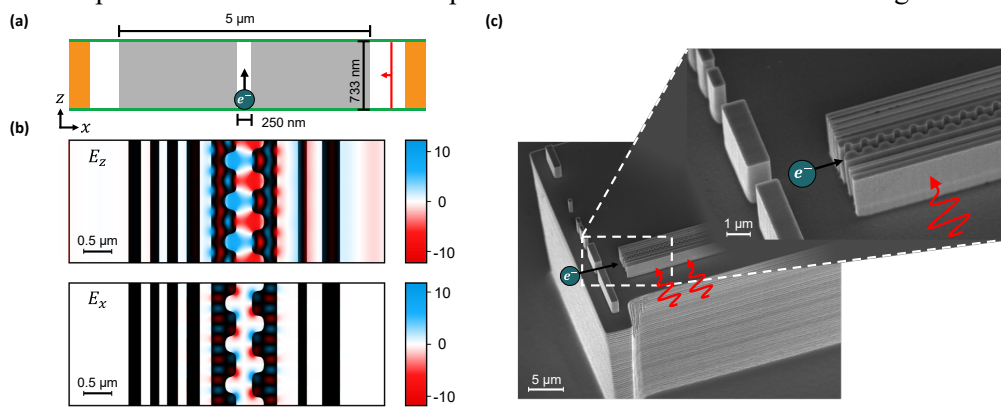

Figure S17: Silicon-photonic nanostructures for efficient electron–light coupling: based on dielectric laser accelerators (DLAs). (a) Schematic illustration of the two-dimensional simulation cell used for inverse design. The nanostructure was optimized over a 5-μm-wide design region (grey) with a 250-nm-broad vacuum channel in the center. Periodic boundaries (green) were applied in longitudinal direction, and perfectly matched layers (orange) were defined at the remaining boundaries. A transverse-magnetic plane wave (red) was excited on one side, and the resulting acceleration gradient, measured in the center of the channel, served as the objective function. (b) Electric field components $E_z$ and $E_x$ normalized by the incoming field amplitude. (c) SEM picture showing our DLA-type nanostructure on a 30-μm-high mesa.